\documentclass[traditabstract]{aa}
\usepackage{graphicx}
\usepackage{epsfig}
\usepackage{color}
\usepackage{txfonts}
\usepackage{natbib}
\bibpunct{(}{)}{;}{a}{}{,} 

\begin{document}

\title{Characteristics of thick disks formed through minor mergers: stellar excesses and scale lengths}
\titlerunning{Minor mergers and thick disk characteristics}

\author{Yan Qu, Paola Di Matteo, Matthew D. Lehnert, \and Wim van Driel}

\authorrunning{Qu et al.}

\institute{GEPI, Observatoire de Paris, CNRS, Universit\'e
Paris Diderot, 5 place Jules Janssen, 92190 Meudon, France\\
\email{yan.qu@obspm.fr}
}

\date{Received, Accepted}

\abstract{By means of a series of N-body/SPH simulations we
    investigate the morphological properties of thick stellar disks
    formed through minor mergers with, e.g. a range of gas-to-stellar
    mass ratios. We show that the vertical surface density profile of
    the post-merger thick disk follows a sech function and has an
    excess in the regions furthest away from the disk mid-plane
    ($z\gtrsim 2$~kpc). This stellar excess also follows a sech
    function with a larger scale height than the main thick disk
    component (except at large radii). It is usually dominated by
    stars from the primary galaxy, but this depends on the orbital
    configuration. Stars in the excess have a rotational velocity
    lower than that of stars in the thick disk, and they may thus be
    confused with stars in the inner galactic halo, which can have a
    similar lag. Confirming previous results, the thick disk scale
    height increases with radius and the rate of its increase is
    smaller for more gas rich primary galaxies. On the contrary, the
    scale height of the stellar excess is independent of both radius
    and gas fraction. We also find that the post-merger thick disk has
    a radial scale length which is $10-50\%$ larger than that of the
    thin disk.  Two consecutive mergers have basically the same
effect on heating the stellar disk as a single merger of the same total
mass, i.e., the disk heating effect of a few consecutive mergers does
not saturate but is cumulative. To investigate how thick disks
    produced through secular processes may differ from those produced
    by minor mergers, we also simulated gravitationally unstable
    gas-rich disks (``clumpy disks'').  These clumpy disks do not
    produce either a stellar excess or a ratio of thick to thin disk
    scale lengths greater than one.  Comparing our simulation results
    with observations of the Milky Way and nearby galaxies shows that
    our results for minor mergers are consistent with observations of
    the ratio of thick to thin disk scale lengths and with the
    ``Toomre diagram'' (the sum in quadrature of the vertical and
    radial versus the rotational kinetic energies) of the Milky
    Way. The simulations of clumpy disks do not show such
    agreement. We conclude that minor mergers are a viable mechanism
    for the creation of galactic thick disks and investigating stars
    at several kpc above the mid-plane of the Milky Way and other
    galaxies may provide a quantitative method for studying the
    (minor) merger history of galaxies.}


\maketitle

\section{Introduction}

Ever since the existence of a thick disk component was shown by star
counts in the Milky Way \citep{gilmoreR183} and by stellar luminosity
distribution studies of external early- and late-type galaxies
\citep{burstein179,tsikoudi179,vdkruitS181a,shawG189,dalcantonB202,
pohlen04,tikhonovGD205}, its origin and properties have been
the subject of a continuing vigorous debate. The distinctive
properties of their stars, e.g. their metallicity distribution
and kinematics, suggest that the thick disk is a distinct and
perhaps intermediate component between the thin disk and the halo
\citep{norris186,wyseG186,sandageF187,carneyLL189,ratnatungaF189}.

A large variety of models has been proposed to explain the formation
of thick disks -- involving either secular heating of the disk through
scattering by spiral waves (\citealt{barbanisW167};\citealt{carlbergS185}),
molecular clouds \citep{spitzerS151,lacey184}, bars, self-gravitating clumps \citep{bournaudEM209},
or heating disk stars through external processes such as galaxy
mergers, infalling clumpy gas coalescing into a thin disk, and
proto-galactic fragments coalescing before thin disk formation.
On the other hand, a thick disk component can also be formed by violent heating and
relaxation due to satellite accretion(s) or from directly
accreting dynamically hot stars through mergers and interactions
\citep{statler188,hernquistQ189,quinnhf193,walker196,aguerri01,
abadiNSE203,yoachimD205,brookRKMG207,villalobosH08,villalobosH09}. As recently discussed
by \citet{salesHAB209} and \cite{dimatteo210}, these different
formation mechanisms should produce different signatures in
the eccentricities of the stellar orbits, thus providing a
potential diagnostic to disentangle the dominant formation scenarios
(\citealp[see][]{dimatteo210,dierickx210,wilson210}, and \citealp[][for
a comparison of model predictions and observations]{casetti210}).
Nevertheless, the main astrophysical processes which drive thick disk
formation remain unclear.

The structure, kinematics and enrichment of the thick disk provide
important clues for solving the mystery of its origin. Many photometric
and spectroscopic observations show that thick disk stars are generally
old, and have lower metallicity and larger velocity dispersion
than thin disk stars \citep{nissen195, chibaB200, gilmoreWN202,
soubiranBS203, parkerHB204}. The vertical surface density profile of
thick disk stars follows a $sech^n$ profile with a scale height
several times larger than the thin disk, whereas their radial
profile has a scale length that is generally 1 to 2 times larger than
that of the thin disk \citep{morrisonMHS197,abeBC199,dalcantonB202,
neeserSMP202,pohlen04,yoachimD205}. In models, the thick disk scale
height is usually assumed to be independent of radius \citep{pohlen04,
yoachimD205}, in accordance with earlier observational studies on thin and thick
disk scale heights \citep{vdkruitS181a,vdkruitS181b, shawG190,
naslundJ197, rauscherLB198}. However, for a sample of 48 early- and
late-type galaxies \cite{degrijs197} found an increase in disk scale
height with radius, where early-type galaxies show the largest increase.
Also \citet{narayanj202b} pointed out that an increasing scale height
was consistent with two of the galaxies studied by \citet{vdkruitS181a,
vdkruitS181b}, and that their vertical velocity dispersion distribution
does not scale as twice the disk scale length with radius, thus supporting
the possibility of a moderate flaring in these disks. Recently,
\cite{bournaudEM209} modeled the formation of a thick disk through
gravitational instabilities and pointed out that thick disks produced
in this way have constant scale heights, while this is not likely to be
the case if they were formed through minor mergers.

Therefore, the properties of vertical disk profiles can help in
understanding their formation processes. It appears that thick disk
scale height is related to both galaxy mass \citep{yoachimD206} and
Hubble type \citep{degrijs197}. Moreover, some disk-dominated galaxies do not show an obvious thick
disk, e.g. NGC 4244 (\citealp{fry199}; \citealp[but see][for evidence of 
a subtle thick disk]{comeron211}), and some
even have thick disks that counter-rotate relative to their thin
disks \citep{morrisonBH194, yoachimD205}.  These observations, taken
all together, favor a merger-induced thick disk formation mechanism
\citep[see, e.g.,][and references therein]{yoachimD205,yoachimD206}.

Interestingly, it has recently been shown by \cite{purcell210}
  and \cite{zolotov210} that minor and major mergers cannot only
  efficiently heat a pre-existing stellar disk, but also contribute to
  the building of galaxy halos. A trace of this heated population may
  be found in the inner Milky Way halo, as recently discussed by
  \cite{nissenS210}.

Despite considerable effort, we are still far from having a detailed
understanding of the origin and properties of galactic thick disks.
In this paper, we investigate if N-body/SPH numerical simulations of
minor mergers, with a mass ratio of 10:1 and 20:1, can produce
realistic thick disks. We concentrate on certain aspects of
post-merger thick disks -- i.e., vertical surface density profiles
and disk scale heights -- that are well suited for comparison with
the observed properties of thick disks in the Milky Way and nearby
galaxies. In particular, we investigate the ``excess'' of stars
at great heights (z$>$2~kpc) from the galaxy mid-plane that is
naturally formed through minor mergers. This stellar excess, as
already discussed by \cite{purcell210} and \cite{zolotov210}, 
may have contributed to the stellar population currently found in
inner galactic halos. For the first time, we describe its main
properties and the impact on them of initial orbital configurations,
primary galaxy gas fraction and consecutive minor mergers. In
addition, we compare the scale length ratios of the thick and the
thin disk to observations, and the Toomre diagrams of minor mergers
with that of the Milky Way. Furthermore, we will compare the
vertical structure of thick disks formed through minor mergers and
through scattering by clumps in the distribution of mass within the
thin disk, and discuss the expected differences in their observable
properties.

The paper is organized as follows. In \S~\ref{model} we describe the
numerical code, the initial galaxy models and orbital conditions
adopted for simulation runs. \S~\ref{results} presents the vertical
structure and kinematics of merger-induced thick disks and their
dependence on initial configuration parameters in our simulations.
Finally, we discuss the results in \S~\ref{discuss} and draw our
conclusions in \S~\ref{conclu}.

\section{Models and initial conditions}\label{model}

We study the interaction and coalescence of a satellite galaxy with
a much more massive disk galaxy. These minor merger simulations are
part of an on-going program to model and understand the role minor
mergers play in the evolution of angular momentum and morphological
and kinematic properties of galaxies \citep[see also][]{quDM210}.
The massive disk galaxy (primary galaxy) consists of a dark matter
halo which is initially not rotating, a central bulge, and a stellar
disk -- hereafter we refer to this model with the nomenclature gS0
(=giant S0 galaxy).  We also consider initial models containing a
gaseous disk, whose mass, $M_{gas}$, is $10\%$ or $20\%$ of that of
the stellar disk, and to which we refer as ``gSa'' and ``gSb'',
respectively.  The spherical dark matter halo and the stellar bulge
are represented by Plummer spheres \citep{binneyT187} and neither
has any rotation at the beginning of the simulation. The stellar and
gaseous disks are represented by Miyamoto-Nagai density profiles
\citep{binneyT187}. Table~\ref{morphtable} lists the total masses of
the halo, bulge and disk components ($M_H$, $M_B$, $M_{*}$ and
$M_{gas}$) and the core radii of the halo and the bulge ($r_H$ and
$r_B$), as well as the vertical and radial exponential scale lengths
of the stellar and gaseous disks ($h _{*}$, $a_{*}$ and $h_{gas}$,
$a_{gas}$). The elliptical satellite galaxy has a total mass equal
to 10$\%$ or 5$\%$ of that of the primary galaxy -- we refer to
these models as dE0 and sE0, respectively. It consists of a
spherical stellar and a dark matter component, both modeled with
initially non-rotating Plummer profiles, whose parameters are also
given in Table~\ref{morphtable}. As we have done in \citet{quDM210},
all galaxy models are evolved in isolation for 1~Gyr before the
interaction starts. The orbital parameters for these minor
interactions have been described in
\citet[][Table~9]{chilingarianDMC209} and \citet[][Table~2]{quDM210}
and we refer the reader to these papers for a detailed description.

In this paper, we follow the nomenclature adopted in \citet{quDM210},
i.e., using a six-character string to indicate the morphology of the
interacting galaxies: the first three describe the type of the primary
galaxy, ``gS0'', ``gSa'' or ``gSb'', depending on its gas fraction,
and the following three describe the satellite galaxy, dE0 or
sE0. This is followed by the suffix ``dir'' or ``ret'', for prograde
or retrograde orbits, respectively. The subsequent two numbers, either
``33'' or ``60'', indicate the initial inclination of the satellite
orbit with respect to the disk plane of the primary galaxy. Thus, for
example, the nomenclature ``gS0dE001dir33'' refers to a prograde
encounter between a gS0 galaxy and a dwarf elliptical satellite
galaxy, with an initial inclination $i=33^{\circ}$, whose initial
orbital parameters are those corresponding to the id=``01dir'' case
listed in Table~9 of \citet{chilingarianDMC209}.

To study the effect of multiple mergers on the properties of
post-merger stellar disk, we ran some simulations in which the
primary galaxy accretes consecutively two satellites over a period
of 3-5~Gyr. We have considered both the case in which both
satellites coming from a fixed direction inclined $33^{\circ}$ with
respect to the primary disk, as well as the case in which the second
satellite comes from a different direction -- hereafter we will
refer to these models as ``multi-A'' and ``multi-B'' respectively.
For example, the nomenclature ``multi-A, 2$\times$10:1'' refers to
the successive mergers of two satellite galaxies whose mass is
one-tenth of that of the primary galaxy and whose orbits are
inclined $33^{\circ}$ with respect to the primary disk.

All simulations (50 in total) were run using the Tree-SPH code
described in \citet{semelinC202}. A total of $N_{TOT}=528,000$
particles have been distributed between the primary and satellite
galaxies (see Table~\ref{numbers}) for all simulations. We also test
the dependence of results on the particle number used in the
simulation by running some additional simulations with a total of
$N_{TOT}=1,056,000$ particles. A Plummer potential is used to soften
gravity on small scales, with constant softening lengths of
$\epsilon=200$ pc ($\epsilon=170$ pc for the high resolution
simulations), for all particles. The equations of motion are
integrated using a leapfrog algorithm with a fixed time step of
0.5~Myr. With these choices, the relative error in the conservation
of the total energy is about $10^{-6}$ per time step.

In this study we also compare the vertical structure of minor
merger-induced thick disks with those generated by an internal
process, through scattering by massive clumps formed in an initially
unstable disk, as recently proposed by \cite{bournaudEM209}. For
this, we analyzed two simulations of gas-rich, unstable disk galaxies
from \citet{dimatteo08}. These use a total number of 120,000
particles, equally distributed among the gas, stars, and dark matter
components, whose parameters are the same as those of the gSb model
(see Table~\ref{morphtable}), except for the gas mass, which is
initially $50\%$ that of the stellar disk. These gas-rich galaxies are
characterized by a very small initial Toomre disk stability parameter
for the gaseous component ($Q_{gas}=0.3$ and $Q_{gas}=0.1$, hereafter
called ``gSb+u1'' and ``gSb+u2'', respectively). Reducing the local
disk stability causes the gaseous disk to fragment into many clumps,
particularly in the ``gSb+u2'' case. Once formed, these clumps migrate
towards the galaxy center; some dissolve under the influence of the
tidal field of the disk, while others survive and merge in the galaxy
center, contributing to the formation of the central bulge. As
recently proposed by \cite{bournaudEM209}, these clumps can scatter
stars in the pre-existing thin disk, contributing to the formation of
a thick disk component.

\begin{table}
\caption[]{Parameters of the initial models of halo, bulge, stellar
and gaseous disks.} \label{morphtable} \centering
\begin{tabular}{lccccc}
\hline\hline & gS0 &  gSa & gSb & dE0 & sE0 \\ \hline
$M_{B}\ [2.3\times 10^9 M_{\odot}]$ & 10. & 10. & 5. & 7. & 3.5\\
$M_{H}\ [2.3\times 10^9 M_{\odot}]$ & 50.  & 50. & 75. & 3. & 1.5\\
$M_{*}\ [2.3\times 10^9 M_{\odot}]$ & 40.  & 40. & 20. & --  & -- \\
$M_{gas}/M_{*}$ & -- & 0.1 & 0.2 & --  & -- \\
$r_{B}\ [\mathrm{kpc}]$ & 2. &  2. &1. & 1.3 & 0.9\\
$r_{H}\ [\mathrm{kpc}]$ & 10. & 10. & 12.& 2.2 & 1.55\\
$a_{*}\ [\mathrm{kpc}]$ & 4. &  4.& 5.& --  & -- \\
$h_{*}\ [\mathrm{kpc}]$ & 0.5 &  0.5 & 0.5& --  & -- \\
$a_{gas}\ [\mathrm{kpc}]$ & --  &  5.& 6.& --  & -- \\
$h_{gas}\ [\mathrm{kpc}]$ & --  &  0.2 & 0.2 & --  & -- \\ \hline
\end{tabular}
\end{table}

\begin{table}
\caption[]{Particle numbers for primary galaxies and satellites}
\label{numbers} \centering
\begin{tabular}{lccccc}
\hline\hline
 & gS0  & gSa & gSb & dE0 & sE0 \\
\hline
$N_{gas}$  & --     &  80,000 & 160,000 & --    & --  \\
$N_{star}$ & 320,000 & 240,000 & 160,000 & 32,000 & 16,000\\
$N_{DM}$   & 160,000 & 160,000 & 160,000 & 16,000 & 8,000 \\
\hline
\end{tabular}
\end{table}

\section{Results}\label{results}

During the merging process, tidal interactions between merging
galaxies deposit part of their energy into stellar motions in the disk
of the primary galaxy, thus causing the disk to become kinematically
hotter and to spread out both radially and vertically, changing its morphology
as well as its kinematics. In this section we will analyze both
dissipationless and dissipative minor mergers and their possible role
in the formation of the thick disk.

\begin{figure*}
\centering
\includegraphics[width=15.cm,angle=0]{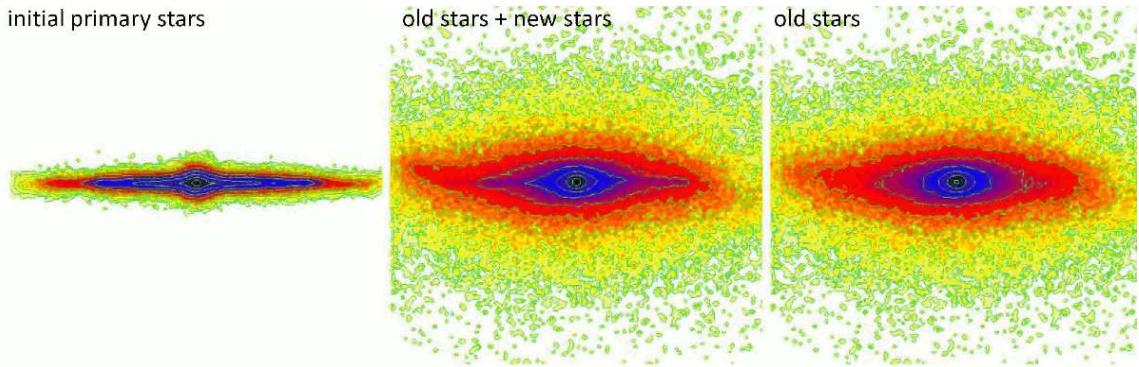}
\vspace{0.2cm} \caption{Stellar distribution in a gSb galaxy model
in its initial configuration (left panel) and at 3~Gyr after the
start of the simulation (middle and right panels). All panels show
the galaxy from an edge-on perspective. Of the post-merger galaxy,
both the original stellar population and the new thin disk formed
during the merger are shown in the middle panel, while in the right
panel we show only stars from the original stellar population of
both galaxies. The thickening of the disk is entirely due to stars
from the original thin disk and the satellite. The same color and
contour scales are used in all panels. Surface density contours are
spaced logarithmically and each panel is 40~kpc$\times$40~kpc in
size.} \label{densitymap}
\end{figure*}

\subsection{Characteristics of the post-merger stellar disk}\label{s-chars}

Here we investigate the vertical stellar distribution of remnant
disks in minor mergers, for which we assume a generalized sech
profile, $\Sigma(r,z)=\Sigma_{r}sech^{2/n}(-\frac{nz}{2z_{0}})$,
where $\Sigma_r$ is the radial exponential surface density profile
and $z_{0}$ is the disk scale height. This profile can fit, for
example, an isothermal sheet if $n=1$, and an exponential if
$n=\infty$ -- the two limiting cases of this generalized profile.
Based on the study of the three-dimensional distribution of disk
stars, \citet{vdkruit188} found that an intermediate disk model with
$n=2$ is the most appropriate \citep[see also][]{schwarzkopfD200},
which is also supported by the theoretical study of
\citet{banerjeeJ207}. We analyze, as far as possible, the properties
of remnant disks as they typically would be observed.  First, we
project the post-merger stellar disk to edge-on, then average the
vertical surface density profiles on both sides of the stellar disk
mid-plane and fit them using a sech profile with $n=2$, i.e.,
$sech(-\frac{z}{z_{0}})$, allowing the scale height parameter
$z_{0}$ to be unconstrained. We also fit the disk scale height for
the two limiting cases of the generalized sech profile, $n=\infty$
and $n=1$: the exponential fit gives an on average 1.4 times larger
disk scale height than the sech profile, whereas the $sech^{2}$ fit
gives a value which is 0.8-0.9 times smaller.

The vertical stellar surface density profiles are analyzed between
0.5 and 1~Gyr after the merger is complete, at a time when the
post-merger stellar disk has settled roughly into a new equilibrium
configuration. At later post-merger times, i.e., $t>1$~Gyr,
the difference in disk scale height with respect to the value at
$t=0.5-1$~Gyr is no more than 15$\%$ inside $r<6r_{d}$. The stellar
disk scale length, $r_{d}$, has been evaluated by projecting the
galaxy to face-on. The difference between the final and initial disk
scale length is always small, $\Delta r_{d}<10\%$, except for
dissipative minor mergers with a gas-to-stellar mass fraction
$f_{gas}=0.2$, where the disk scale length shrinks by $\sim
15-20\%$.  Thus the exact time at which the merger simulations are
analyzed beyond 0.5-1~Gyr after the completion of the merger makes
little difference in the final results. In this study we
also mask the contribution of the bulge component by excluding stars that were
initially in the bulge of the primary galaxy and we consider only
stars which were initially in the disk of the primary galaxy or in
the satellite.

We note that in the case of dissipative minor merger
simulations, a new stellar disk forms during the merging process due
to star formation in the gaseous disk. This ``new'' stellar disk
component is thin, with scale heights of $300-400$ pc only (see
Fig.~\ref{densitymap} as an example). We will not discuss in detail this
newly formed thin disk component, but concentrate on the old stellar
disk, i.e., the one that was already in place before the interaction,
in order to make a direct comparison with that found in
dissipationless mergers.  We do not find any young stars at scale
heights larger than that of the thin disk formed during the merger.

In the rest of this section, we will examine the vertical structure
of the remnant disk after a single dissipationless minor merger,
to illustrate the process of thick disk formation during a gas-free
minor merger. We will present the vertical stellar distribution and the
fitting results, and will later compare these properties with those
of remnant thick disks found in dissipative or two consecutive
minor merger simulations, as well as with thick disks formed
through massive clumps in a gravitationally unstable disk.

\subsubsection{Vertical structure of the post-merger thick
stellar disk: evidence for a stellar excess}

\begin{figure}
\centering
\includegraphics[width=7.8cm,angle=0]{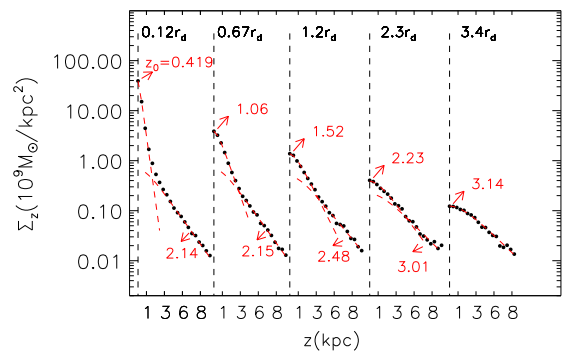}
\vspace{0.cm} \caption{Vertical stellar surface density profiles of
a post-merger thick stellar disk at 0.5 Gyr after the completion of
a dissipationless minor merger (black). We show profiles at a
variety of distances from the galaxy center along the galactic
mid-plane, in units of the disk scale length $r_d$. Also shown are
the double sech functions fitted to the profiles and their
corresponding scale heights $z_0$ in kpc (labeled in red).}
\label{dens-fit}
\end{figure}

The vertical surface density profile of a thick disk formed
  through a dissipationless minor merger is shown in
  Fig.~\ref{dens-fit}. This profile has been evaluated 0.5~Gyr after
  the completion of the merger. The main features of the profile fits are:

\begin{itemize}
\item {A single sech function cannot fit the entire vertical profile.
While the lower regions, at z$\le$2~kpc, can be described by a sech function,
after removing this sech function there remains a large excess of stars
at greater heights above the mid-plane (z$>$2~kpc);}

\item This excess is especially prominent in the inner regions of the
  disk, at $r\lesssim 3r_{d}$, and it virtually disappears at larger radii;

\item The excess has a scale height several times larger
  than that of the main thick disk component (i.e., the one at z$\le$2~kpc):
at $r=0$, for example, the scale height of the stellar
  excess is $z_{0}\approx 2.3$~kpc, about 5 times that of
  the main thick disk component, which has $z_{0}\approx 0.42$~kpc;

\item {The excess has a scale height that is approximately
  constant with distance from the galaxy center, while the thick disk
scale height increases with radius, as already shown by \cite{bournaudEM209} and
\cite{kazantzidis208} -- this will be discussed further in \S~\ref{depend};}

\item {Due to this excess, two sech functions of different
  scale heights are needed to fit the entire vertical stellar
  profile.}
\end{itemize}

\subsubsection{Thick disk and stellar excess: formation, growth and composition}

\begin{figure}
\centering
\includegraphics[width=7.3cm,angle=0]{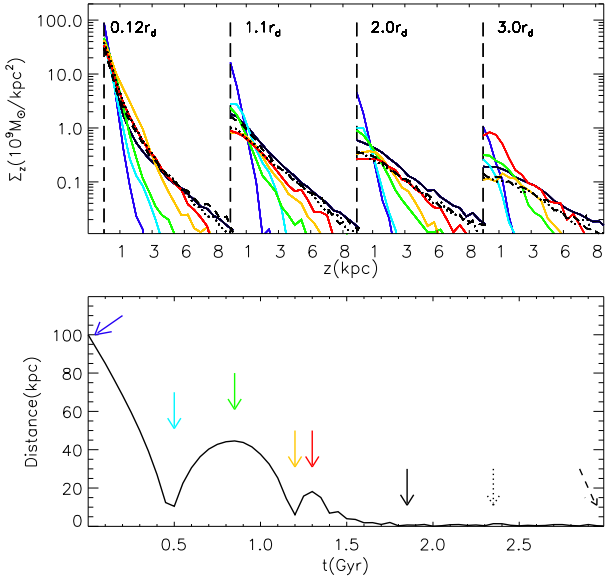}
\vspace{0.4cm} \caption{{\it Upper panel}: Vertical stellar surface
density profiles as function of radius and time (indicated by color; see lower panel) for
a dissipationless minor merger.  The radius is indicated by the 4 black labels.
{\it Lower panel}: The distance between the centers of the satellite and
the primary galaxy as function of time. Arrows mark the merger phases which correspond to
profiles with the same color and line styles shown in the upper
panel.} \label{excess}
\end{figure}

How do the thick disk and the stellar excess develop in the
simulations? We show an example of the formation and growth of a thick disk
in Fig.~\ref{excess}. After the first pericenter passage of the satellite
galaxy ( at $t=0.5$~Gyr) the thin stellar disk component has already
thickened and grown vertically, and after the second pericenter passage, at
$t=1.5$~Gyr, the stellar disk has acquired the double sech profile.
Since the stellar excess is already in place well before the satellite
merges with the primary it is not associated generally with stars from
the satellite galaxy, but rather composed of stars which were
initially in the primary disk and subsequently heated, reaching greater
heights during the merging process. In all
dissipative mergers and in most of dissipationless ones, stars initially
in the disk dominate at all disk heights, in agreement with the
results of \cite{purcell210}. Only in few dissipationless mergers do
we find that the outer vertical profile is dominated by stars from the
satellite galaxy.

The formation of a double sech profile in the vertical density
distribution is not due to the presence of the bulge. Although we
excluded bulge stars from the analysis of the vertical profile of the
thick disk, we did run simulations of dissipationless minor mergers
with bulgeless galaxies. In these simulations we also found a stellar
excess with characteristics very similar to simulations including a
bulge component.

In summary, stars in the stellar excess have their origin in
  different regions of the thin disk, which means that they are not
  only heated vertically, but also redistributed radially. As
  discussed in \S~\ref{discuss}, this has consequences for the expected
  metallicity gradient in the merger remnant.

\subsection{Varying orbital and morphological parameters}\label{depend}

\subsubsection{Dependence on initial orbital configurations}
Regardless of the initial orbital parameters (orbital energy and
angular momentum, direct and retrograde orbits, initial orbital
inclination of the satellite orbit) we always find a thickened
stellar disk in the merger remnant, whose scale height increases
with the radial distance along the major axis of the disk-plane
(Fig.~\ref{thick}). For the dissipationless mergers over the
range $r_{d}<r<3r_{d}$ the scale height is $z_{0}=1.4-2.5$~kpc, or
about 3-6 times larger than the $z_{0}$ value near the disk center,
where $z_{0}=0.4$~kpc. The six sets of different orbital
configurations produce a scatter around the mean value of $|\Delta
z_{0}|/z_{0,mean}\lesssim15\%$. The increase in the disk scale
height with radius implies that the disk heating is the most
significant in the outer regions of the disk. We do not find
such a strong thickening of the stellar disk if the galaxy is evolved
in isolation for 3~Gyr (see Fig.~\ref{thick}): the disk scale height
increases only marginally, less than 25$\%$ at $r\sim r_d$,
and the thickening is independent of radius.

Early research in self-gravitating stellar disks suggested that the
thick disk scale height is independent of radius, assuming that the
vertical velocity dispersion decreases exponentially with radius and
relates to the disk scale length as $h_{vel}=2r_{d}$
\citep{vdkruitS181a}. \citet{narayanj202b} generalized this picture
by taking into account the gravitational contributions from the gas
and the dark matter halo and allowing the ratio $h_{vel}/r_{d}$ to
vary. Contrary to previous results, they found that scale heights
increased with disk radius within 4 or 5 disk scale lengths, and that
the $h_{vel}/r_{d}$ ratio ranged from about 2 to 4.

As pointed out by \citet{quinnhf193}, in mergers of primary galaxies
with their satellites, the stellar disk is heated in three dimensions
and although the velocity dispersion profiles of the remnant disk are
similar to those of the initial disk, they show more extended
exponential profiles. Fitting both the initial and post-merger
vertical velocity dispersion profiles with exponential functions,
we found that while the initial profiles usually have
$h_{vel}/r_{d}\sim 2$, in agreement with an initial stellar disk of
constant scale height, as assumed in our model, the $h_{vel}/r_{d}$
ratio of the post-merger vertical velocity profile was increased to the
range of $\sim 2.5-4$, which is in agreement with what \cite{narayanj202b} found.

From the definition of $z_{0}$ deduced from Poisson's and Jeans'
equations, this link between the $h_{vel}/r_{d}$ ratio and the thick
disk scale height is not difficult to understand. Assuming the
velocity dispersion follows an exponential profile, the definition of
$z_{0}$ \citep[see Eq.(3) in][]{vdkruitS181b} can be rewritten as

\begin{eqnarray}
z_{0}=C_{0}\exp\left[-\left(\frac{2r_{d}}{h_{vel}}-1\right)\frac{r}{2r_{d}}\right]
\end{eqnarray}

where $C_{0}=\frac{\sigma_z(r=0)}{\rho_{0}(2\pi G)^{1/2}}$ and
G is the gravitation constant. $\sigma_z(r=0)$ and $\rho_{0}$ are
the vertical velocity dispersion and the volume density at the
galactic center, respectively. It is quite clear from Eq.(1) that
$z_{0}$ is independent of radius if $h_{vel}/r_{d}=2$, whereas
$z_{0}$ increases with radius when $h_{vel}/r_{d}>2$.

Fig.~\ref{thick} also shows the scale height of the stellar excess
present in the vertical surface density profiles seen at $z\gtrsim 4-5$
kpc and $r\lesssim 3r_{d}$ (see also Fig.~\ref{dens-fit}).  The final
scale height of the stellar excesses in the simulations are independent
of radius and are sensitive to the characteristics of the satellite
orbit. Variations in the orbital parameters lead to a large scatter
in the vertical scale height of the stellar excesses, of up to 30$\%$
($|\Delta z_{0}|/z_{0,mean}\lesssim30\%$).

\subsubsection{Dependence on satellite central density}

\begin{figure}
\centering
\includegraphics[width=5.7cm,angle=0]{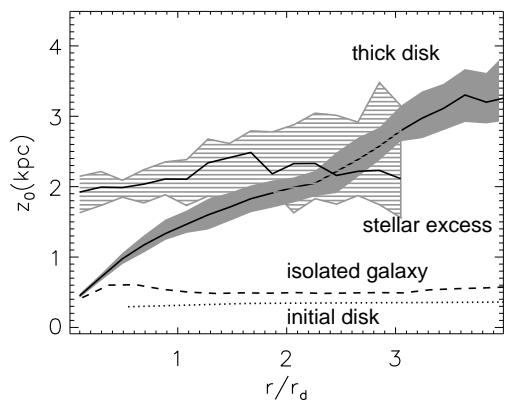}
\vspace{0.8cm}
\caption{The scale heights $z_0$ of the merger-induced
thick disk and stellar excess as function of radius in units of disk scale length,
$r/r_{d}$, for dissipationless minor mergers with six sets of
different initial orbital parameters (see also Fig.\ref{dens-fit}).
The shaded regions indicate the scatter in the scale heights. Most
of this scatter results from the differences in initial orbital
configurations. Also shown are the scale heights of the original
stellar disk (dotted line) and after its evolution in isolation for
3~Gyr (dashed line).}
\label{thick}
\end{figure}

\begin{figure}
\centering
\includegraphics[width=5.7cm,angle=0]{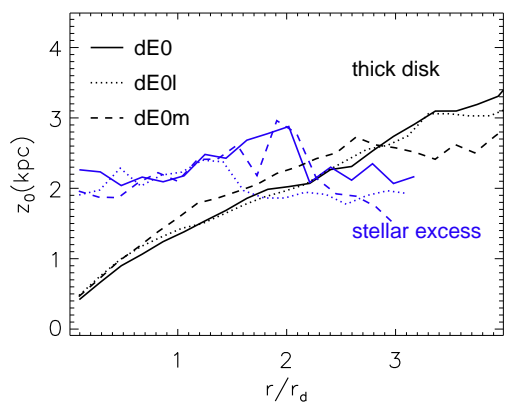}
\vspace{0.8cm}
\caption{The scale heights of the merger-induced thick
disk and stellar excess as function of radial distance in units of disk scale length,
$r/r_{d}$, for dissipationless minor mergers with three different baryonic
central mass concentrations of the merging satellite. The mass of the satellite is the same,
but the central volume density is 50$\%$ higher (dE0m, dashed lines) or 50$\%$
lower (dE0l, dotted lines) than in the reference satellite dE0 (solid lines).}
\label{thick-density}
\end{figure}

Do the flaring of the thick disk and the scale height of the
stellar excess found in Fig.~\ref{thick} depend on the central density
of the satellite galaxy? To investigate
this point, we ran some simulations changing the central density of the
baryonic component of the dE0 satellite, using satellites having
a central volume density $50\%$ higher (dE0h) or lower (dE0l) than
the reference dE0 galaxy. A denser satellite heats the
thick disk more effectively, especially in the inner regions of the primary galaxy (inside
$\sim 2r_d$), thus resulting in a thick disk scale height about $20\%$
higher than that found after the merger of less dense satellites. However,
no clear dependence on the satellite central density is found for the
characteristics of the stellar excess (see Fig.~\ref{thick-density}).

\subsubsection{Dependence on gas fraction}

\begin{figure}
\vspace{0.cm}\hspace{0.5cm}
\centering
\includegraphics[width=8.2cm,angle=0]{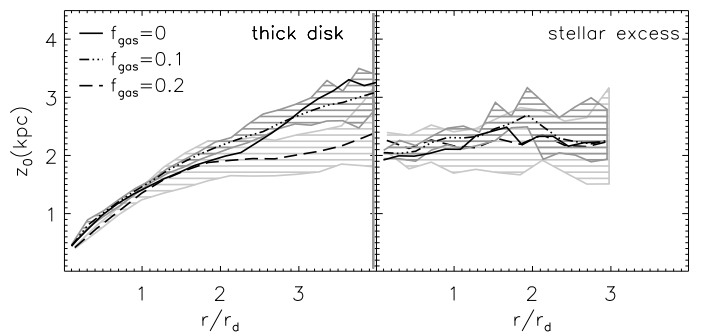}
\vspace{0.8cm} \caption{Scale heights of the merger-induced
thick disk and stellar excess as function of radial distance in units of
disk scale length, $r/r_{d}$, for minor mergers with three different
gas-to-stellar mass fractions, $f_{gas}=0$, 0.1 and 0.2. The shaded regions
indicate the scatter in the scale heights due to the variety of
initial orbital configurations.} \label{thick-gas}
\end{figure}

How does including gas in the primary galaxy's disk affect the outcome
of these simulations?  In Fig.~\ref{thick-gas} we compare the disk
scale heights of dissipative minor mergers (with gas-to-stellar mass
fractions $f_{gas}=0.1$ and $0.2$) with those of dissipationless
mergers. A thick stellar disk is present in all merger
remnants, independent of the gas fraction in the primary, and its
scale height always increases with radius. Dissipative mergers with
$f_{gas}=0.1$ have similar scale heights as the dissipationless
(gas-free) ones. However, in mergers with $f_{gas}=0.2$ a
significantly thinner remnant stellar thick disk is formed, of which
the scale height at $r>2r_{d}$ is on average $\sim$20$\%$-30$\%$ smaller
than in dissipationless mergers. Interestingly,
this reduction is comparable to the 25$\%$ found for 10:1 dissipative
mergers with 20$\%$ gas fraction by \citet{mosterMS209}, which
indicates that their conclusions are robust to the use of different initial
galaxy models, orbital conditions and numerical codes.

On the other hand, the stellar excess in dissipative minor
mergers still has scale heights that are independent of radius, and very
similar to those found in dissipationless mergers (see left panel in
Fig.~\ref{thick-gas}). This implies that the dissipative gas component
and further induced star formation mainly affects the higher density
regions close to the remnant disk mid-plane and that it has almost no
effect in regions at greater heights above the disk mid-plane.  This
is consistent with the excess being predominately due to tidal effects
early in the merger event.

\subsection{Dissipative minor mergers versus clump instabilities}\label{gas}

\begin{figure}
   \centering
\includegraphics[width=7.5cm,angle=0]{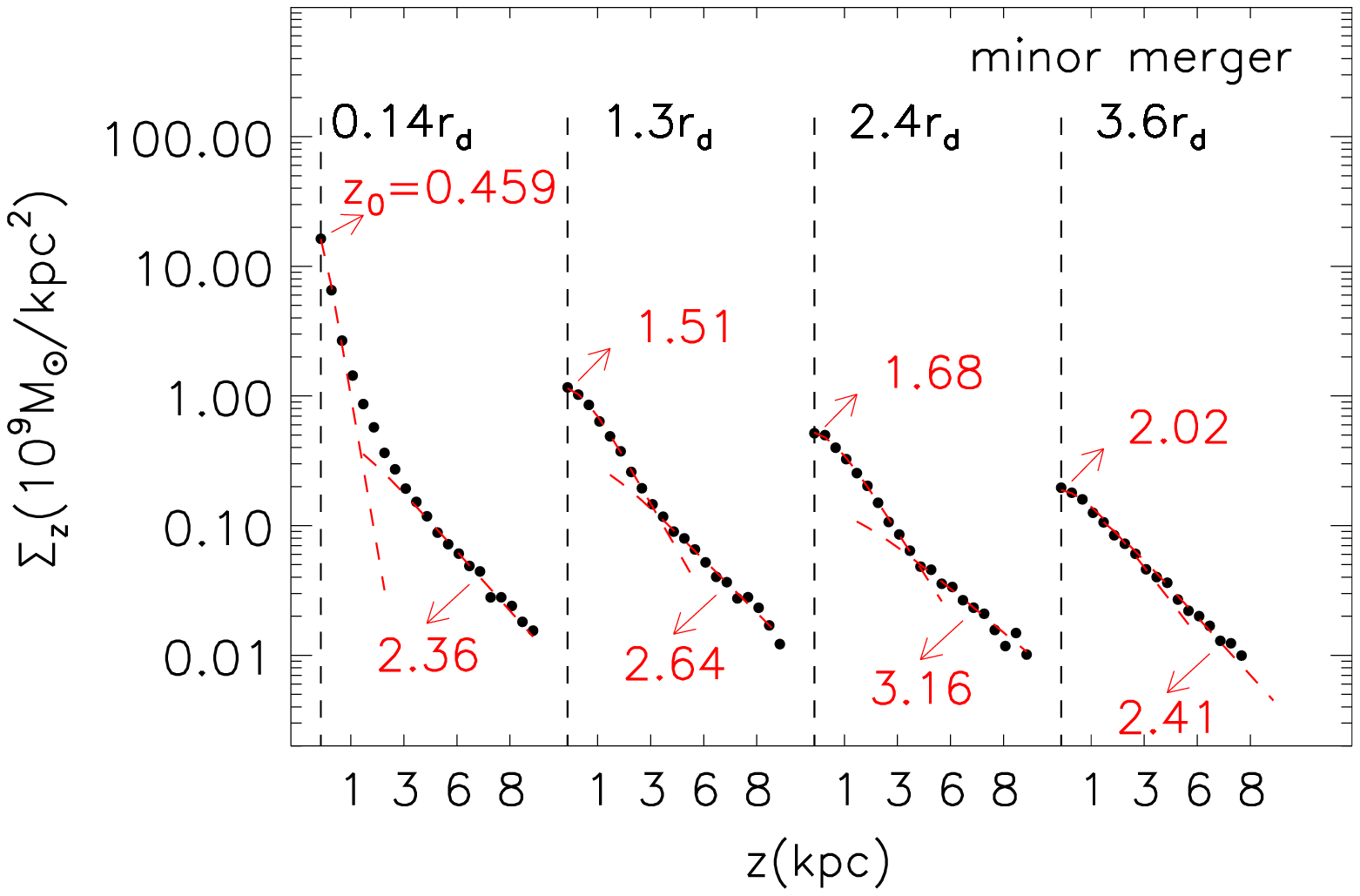}
\includegraphics[width=7.5cm,angle=0]{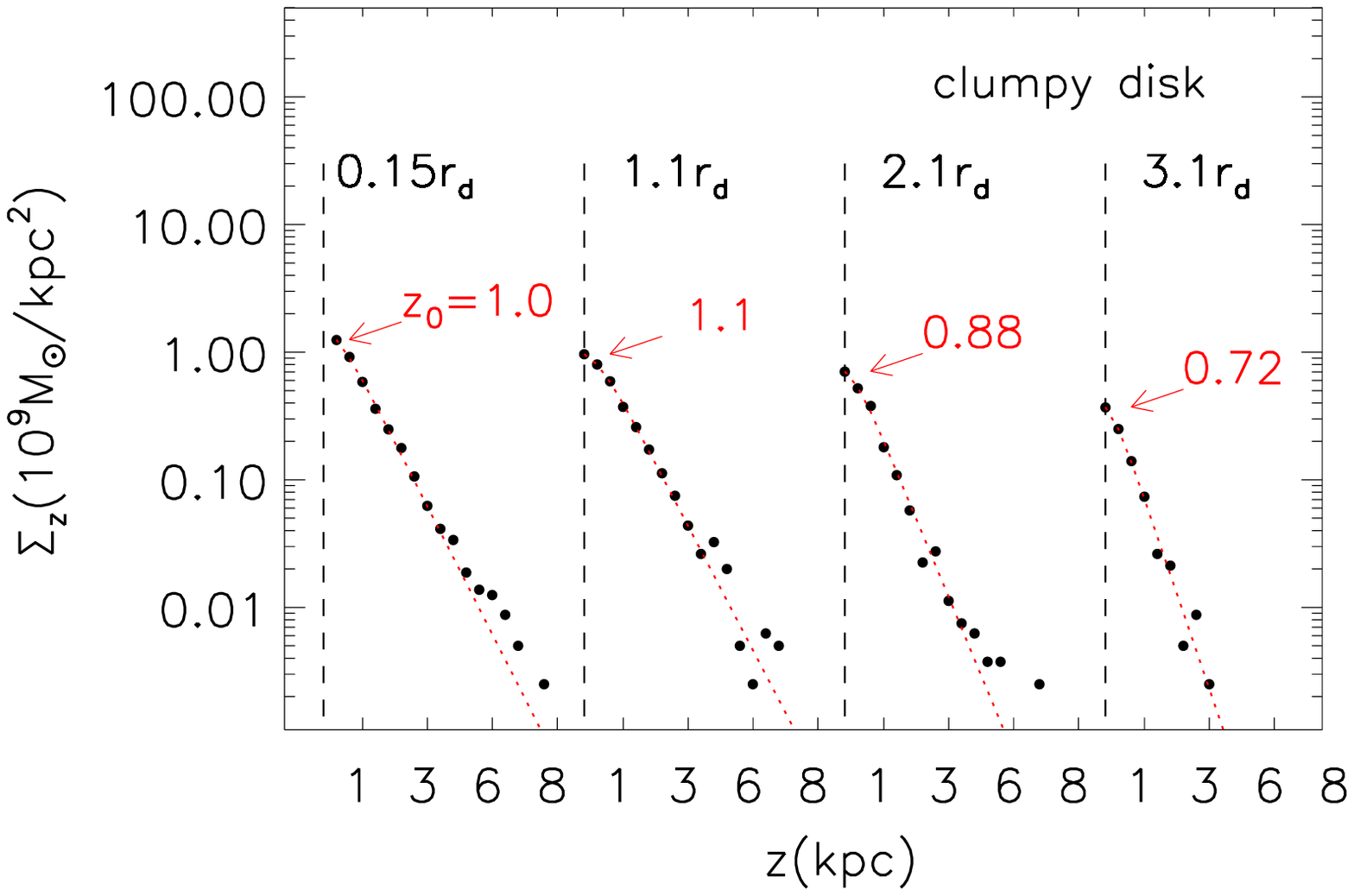}
\caption{Vertical density profiles of the post-merger stellar disk
in a dissipative minor merger with $f_{gas}=0.2$ (upper panel) and of an
isolated ``clumpy disk'' galaxy with $f_{gas}=0.5$ (lower panel). Profiles
are shown at four different exponential disk scale lengths, $r_d$,
from the galaxy center along the major axis of the galactic mid-plane.
The best fitting sech(z) functions to these profiles and corresponding
scale heights are also shown (in red).}
\label{dens-gas}
\end{figure}

It is of great interest to investigate the differences in the
properties of merger-induced thick disks to those formed by secular
processes such as scattering due to clumps formed in a
gravitationally unstable disk, as recently suggested by
\cite{bournaudEM209}. To this end, we compare the vertical surface
density profiles of thick disks induced by minor mergers with those
formed in gas-rich galaxies which are initially unstable to clump
formation and were evolved in isolation for 3~Gyr (see
\S~\ref{model} for details of these galaxy models).  We found that
the resulting scale height in the clumpy isolated galaxies is about
2 times larger, $z_{0}\approx 0.85$, than the initial value and
nearly constant with radius, thus confirming the results of
\citet{bournaudEM209}. Scattering by gas clumps do not only
produce thick disks with constant scale heights, but also vertical surface density
  profiles which do not show any stellar excess (Fig.~\ref{dens-gas}).
In principle, this may indicate an observational way to
distinguish a secular from a merger origin for thick disks.  We
will elaborate on this further in the \S~\ref{discuss}.

\subsection{Consecutive minor mergers and their impact on the thick disk
and stellar excess scale heights}\label{multi}

In hierarchical structure formation scenarios, galaxies grow through
both minor and major mergers and periods of gas accretion from either
cooling gas in their halos or from cosmic filaments. Satellite accretion
is expected to be quite common in a $\Lambda$CDM universe and important
for galaxy growth \citep{guo08}. The dynamical timescale for a satellite
in the dark matter halo of a significantly more massive primary galaxy
can be, and often is, long compared to the merging time scale. In such a
situation we may expect multiple minor mergers to be common since this
allows the primary galaxy to have several satellite galaxies that have
yet to merge.

What is the effect of consecutive mergers on the vertical
stellar distribution of disks and their impact on the thick disk and
stellar excess? Do the effects ``saturate'' -- in other words, does the
additional heating caused by each subsequent merger have an increasingly
smaller effect on heating of the disk? In order to gain insight in this,
we compared the impact on the heating of the disk of a single minor merger
to that of two sequential minor mergers. We adopted four merger histories
and two mass-ratios -- a single 20:1 merger and two sequential 20:1 mergers,
a single 10:1 and two sequential 10:1 mergers, which we will label as
20:1, 2$\times$20:1, 10:1, and 2$\times$10:1, respectively. All of these
mergers are dissipationless and intended only as an initial exploration
of the effects of multiple minor mergers on galaxy disks.

\begin{figure}
\centering
\includegraphics[width=7.5cm,angle=0]{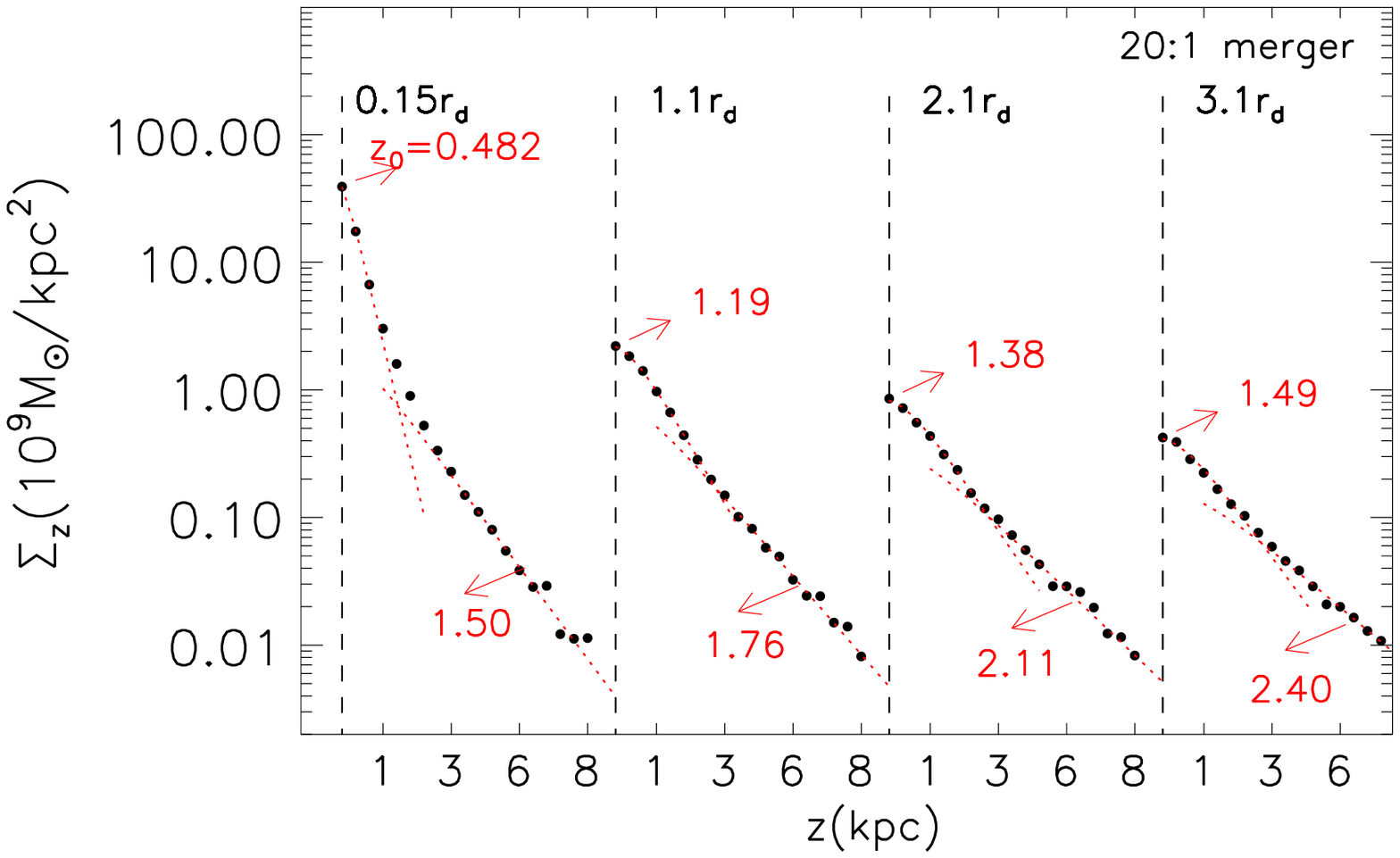}
\vspace{-0.5cm}
\includegraphics[width=7.5cm,angle=0]{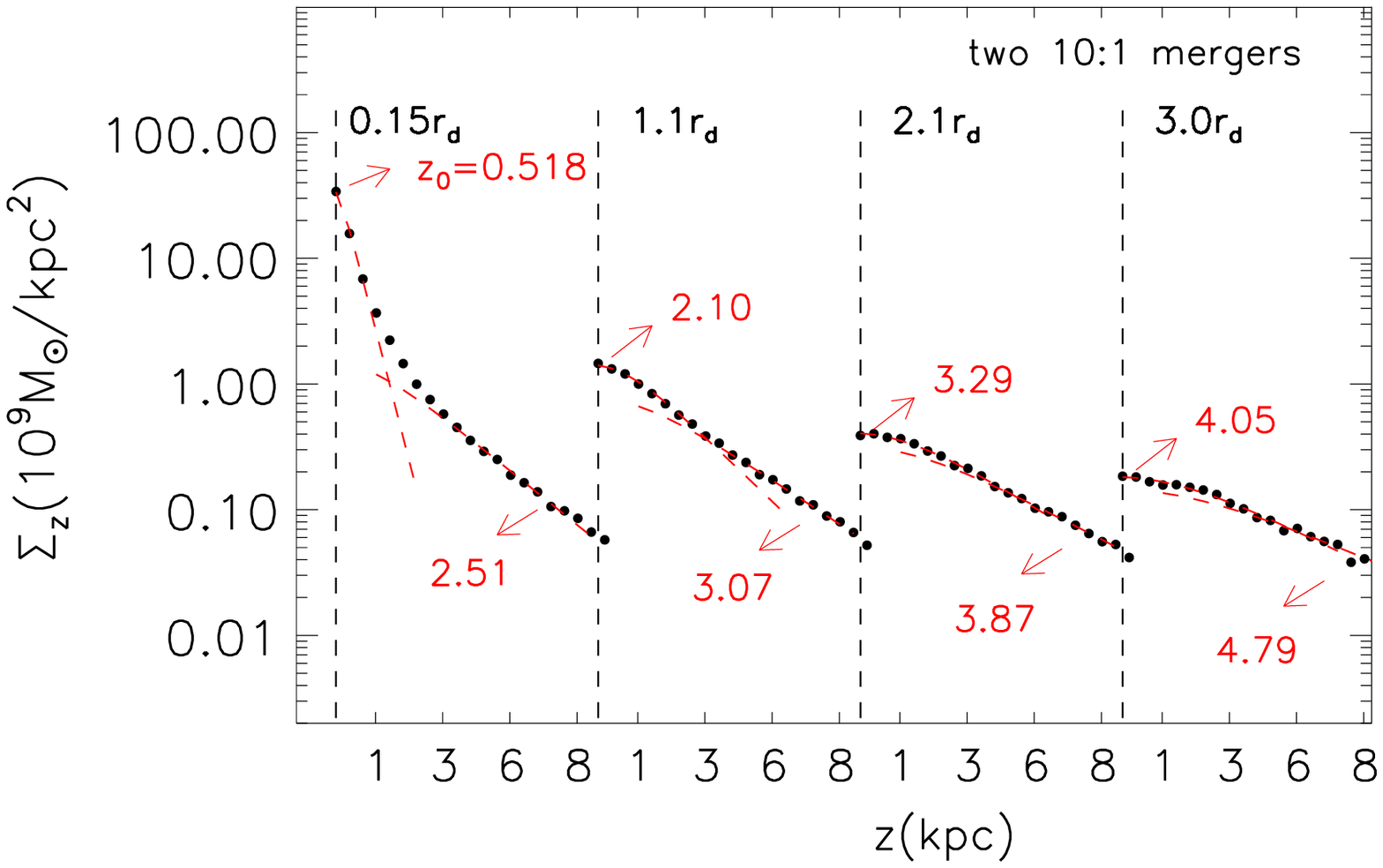}
\vspace{0.cm} \caption{Vertical stellar surface density
profiles of a post-merger thick stellar disk 0.5~Gyr after the
completion of a single dissipationless minor merger with a mass ratio of
20:1 (upper panel), and two consecutive dissipationless minor
mergers with a mass ratio of 10:1 (lower panel). We show profiles
at four different exponential disk scale lengths, $r_d$, from the
galaxy center along the major axis of the galactic mid-plane.
Superposed on these distributions are the fitted double sech functions
and their corresponding scale heights (in red).}
\label{dens-fit-multi}
\end{figure}

\begin{figure}
\vspace{0.cm}\hspace{0.5cm}
\centering
\includegraphics[width=8.2cm,angle=0]{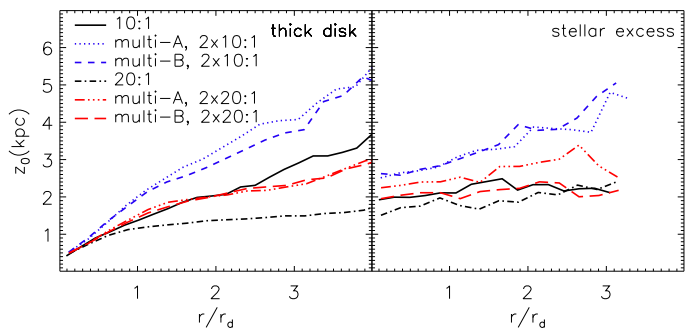}
\vspace{0.8cm} \caption{Scale heights of the merger-induced thick disk
and stellar excess as function of radial distance in units of disk scale length,
$r/r_d$, after a single or two consecutive minor mergers. We show single
mergers with a mass ratio of 10:1 and 20:1, and 2$\times$10:1 and 2$\times$20:1
consecutive merger events. In the consecutive merger cases the satellites
are either accreted from the same direction (multi-A) or from different directions (multi-B).}
\label{thick-mm}
\end{figure}

The vertical surface density profiles of the remnant of a 20:1 and
a 2$\times$10:1 merger are shown in Fig.~\ref{dens-fit-multi}.  Comparing
these profiles with those of a single 10:1 merger (Fig.~\ref{dens-fit}),
we find, as expected, that the thickness of the final disk increases with
the total mass acquired by a galaxy through mergers. In other words,
the more mass accreted, the more internal energy accrued, and the more
efficient the heating of the stellar disk. But changing the mass of the satellite has an
impact also on the formation and resultant scale height of the stellar
excess. A 20:1 minor merger bring less energy to the system, which
means not only a smaller thick disk scale height, but also a obscure stellar excess: stars originally
in the thin disk are not heated enough to reach high vertical distances
from the galaxy mid-plane. On the other hand, a 2$\times$10:1 heats the
thick disk considerably, especially in the outer parts. This depends on
our original definition of thick disk and stellar excess. The behavior
of thick disk and stellar excess scale heights as a function of radius,
for six representative single and two consecutive mergers, is shown in Fig.~\ref{thick-mm}.

In the outer regions, beyond $2r_{d}$, the scale height of the remnant
disk in the single 10:1 minor merger case is $\sim$1.5 times larger
than in the 20:1 merger, whereas the 2$\times$20:1 merger has a similar disk scale
height as a single 10:1 minor merger. Varying the orbital inclination
of the second satellite (``multi-A'' and ``multi-B'' cases) only causes
a rather trivial change in the disk scale height. The effect on disk
scale height of two 20:1 mergers appears to be cumulative, and almost the
same as for a single 10:1 merger (Fig.~\ref{thick-mm}).  Thus comparing
a 2$\times$10:1 merger is like comparing with a 4$\times$20:1 merger,
provided the effect of multiple mergers does not saturate. From the very
simple arguments of Liouville's theorem and the conservation of energy, this is
expected to be the case for dissipationless systems \citep{binneyT187}.

In our simulations of a 2$\times$10:1 minor merger, after the first
10:1 merger the resultant thick disk still responds dynamically to the
heating induced by the subsequent 10:1 merger (Fig.~\ref{thick-mm}).
After the second merger the final disk scale height is $1.5-2$ times
larger than in a single 10:1 merger. Similarly, the stellar excesses
found at greater heights in the vertical stellar distribution have
also increased in the subsequent merger events (Fig.~\ref{thick-mm})
-- their disk scale heights after a single 10:1 merger are 1.5 times
larger than that after a single 20:1 merger. In the 2$\times$10:1 mergers,
their scale heights are not only $\sim 1.4-2$ times larger than in
single 10:1 mergers but they also increase with radius.  Thus, we find that
the effect of two consecutive minor mergers in dissipationless systems
does not saturate but is cumulative. However, we note that after the
second 1:10 merger the stellar excess begins to develop a scale height that depends (weakly)
on radius, and is more pronounced at larger radii where its scale height becomes
very similar to that of the thick disk itself (Fig.~\ref{thick-mm}).  Thus,
at least for the highest mass ratios and in the outer disk, the stellar
excess becomes almost indistinguishable from the thick disk.
For most of our two consecutive interaction models we confirm what we
found for the single mergers, that is stars from the primary disk galaxy
dominate the vertical stellar distribution everywhere in the thick disk.

\subsection{Kinematics of thick disk and stellar excess}\label{kinem}

\begin{figure}
\centering
\includegraphics[width=8.2cm,angle=0]{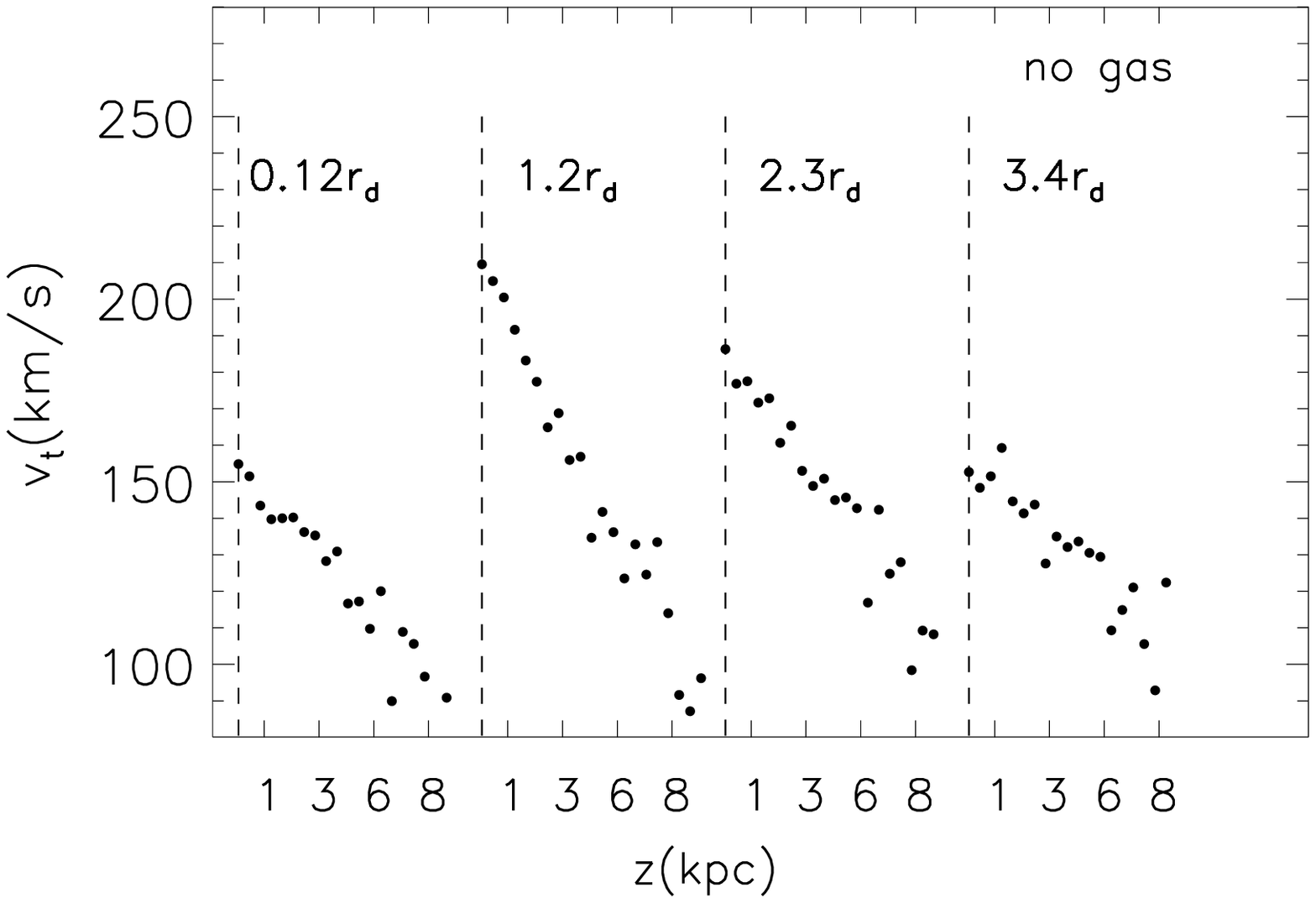}
\includegraphics[width=8.2cm,angle=0]{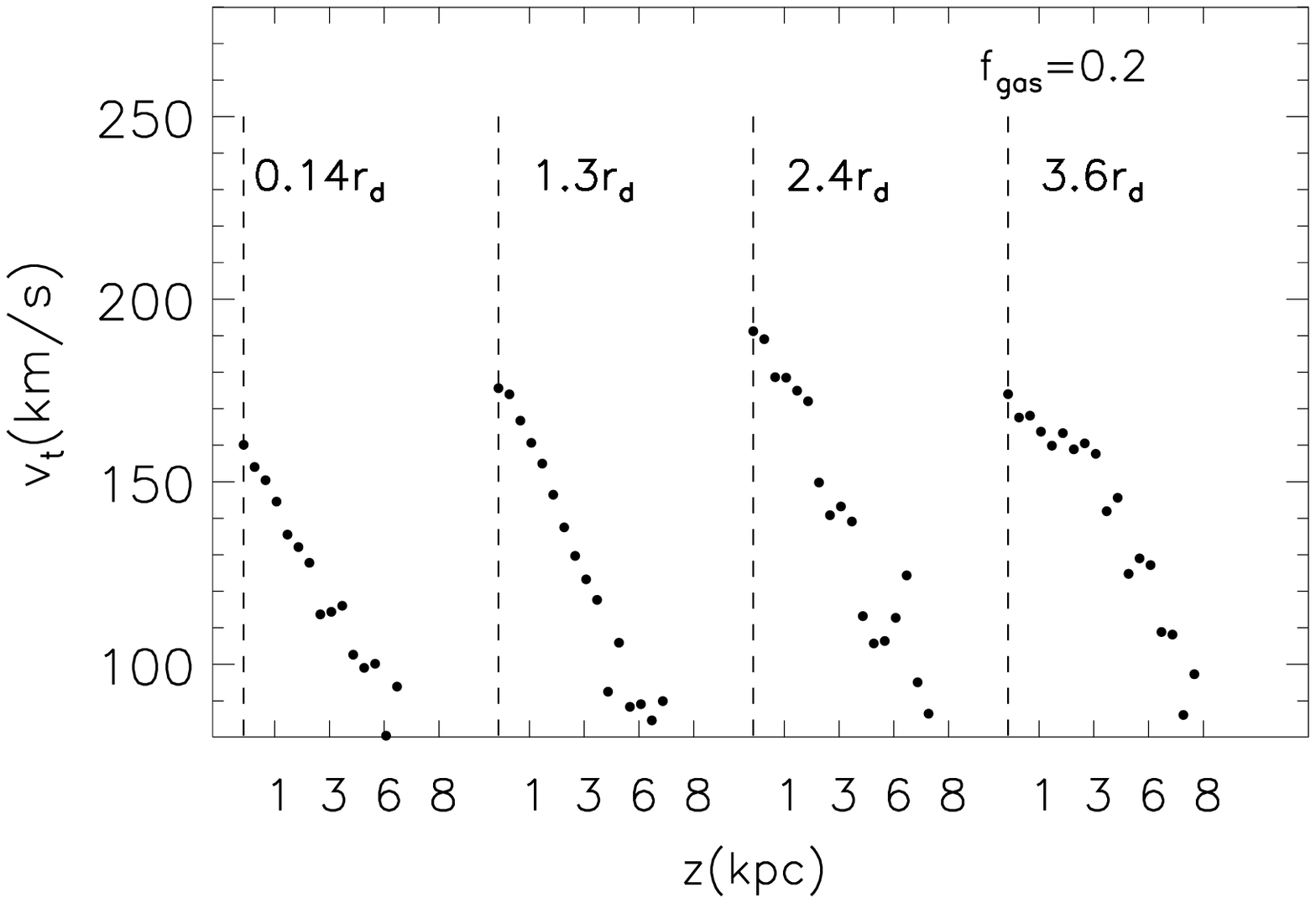}
\caption{Tangential velocity profiles of the post-merger stellar
disk as function of height from the galaxy mid-plane in
dissipationless (upper panels) and dissipational (lower panels)
minor mergers with a gas mass fraction of 0.2. Profiles are shown at
different disk scale lengths, $r_d$, from the galaxy center along
the major axis of the galactic mid-plane.} \label{lag}
\end{figure}

Do the thick disk and stellar excess have similar kinematic
properties? A detailed study of various kinematic properties of minor
merger remnants will be made separately \citep{quDM211},
and here we will discuss only one of the kinematical characteristics
of the stellar excess: its rotational lag with respect
to thick disk stars. This rotational lag is clear in Fig.~\ref{lag},
where the tangential velocity of stars is shown as a function of the
vertical distance from the galaxy mid-plane, for a dissipationless and
a dissipative merger. In both cases the tangential velocity decreases
with distance from the mid-plane, thus indicating that stars far from
the mid-plane plane, which populate the stellar excess, have lower specific
angular momentum than stars in the thick disk (which are closer
to the mid-plane). This is consistent with the trend found
by \cite{dimatteo210} for the eccentricities of stellar orbits after
minor mergers: stars tend to move on more eccentric orbits as their
vertical distance increases and thus tend to rotate more slowly than
they did originally.

This increase in the rotational lag is qualitatively in agreement with
observations \citep{ivezic208}, and based on our models we would
expect to find high metallicity, small lag stars at low height, and stars
with lower metallicity and larger lag at greater height -- this is due to
the way the thick disk is formed in our model, through the heating of
a thin disk that continuously forms stars.

\section{Discussion}\label{discuss}

\begin{figure*}
\centering
\vspace{0.3cm}\hspace{0.7cm}
\includegraphics[width=4.8cm,angle=0]{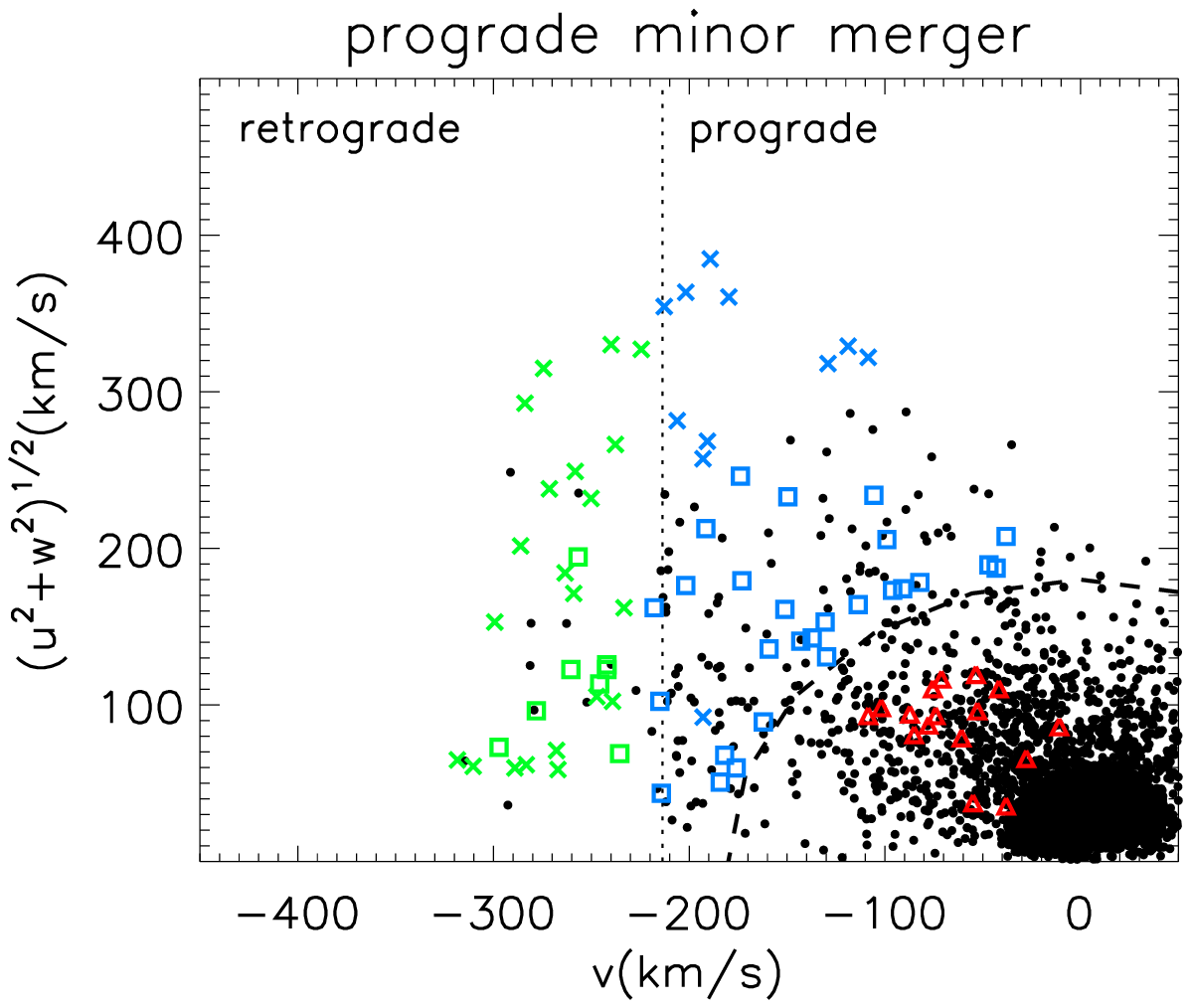}
\vspace{0cm}\hspace{1.2cm}
\includegraphics[width=4.8cm,angle=0]{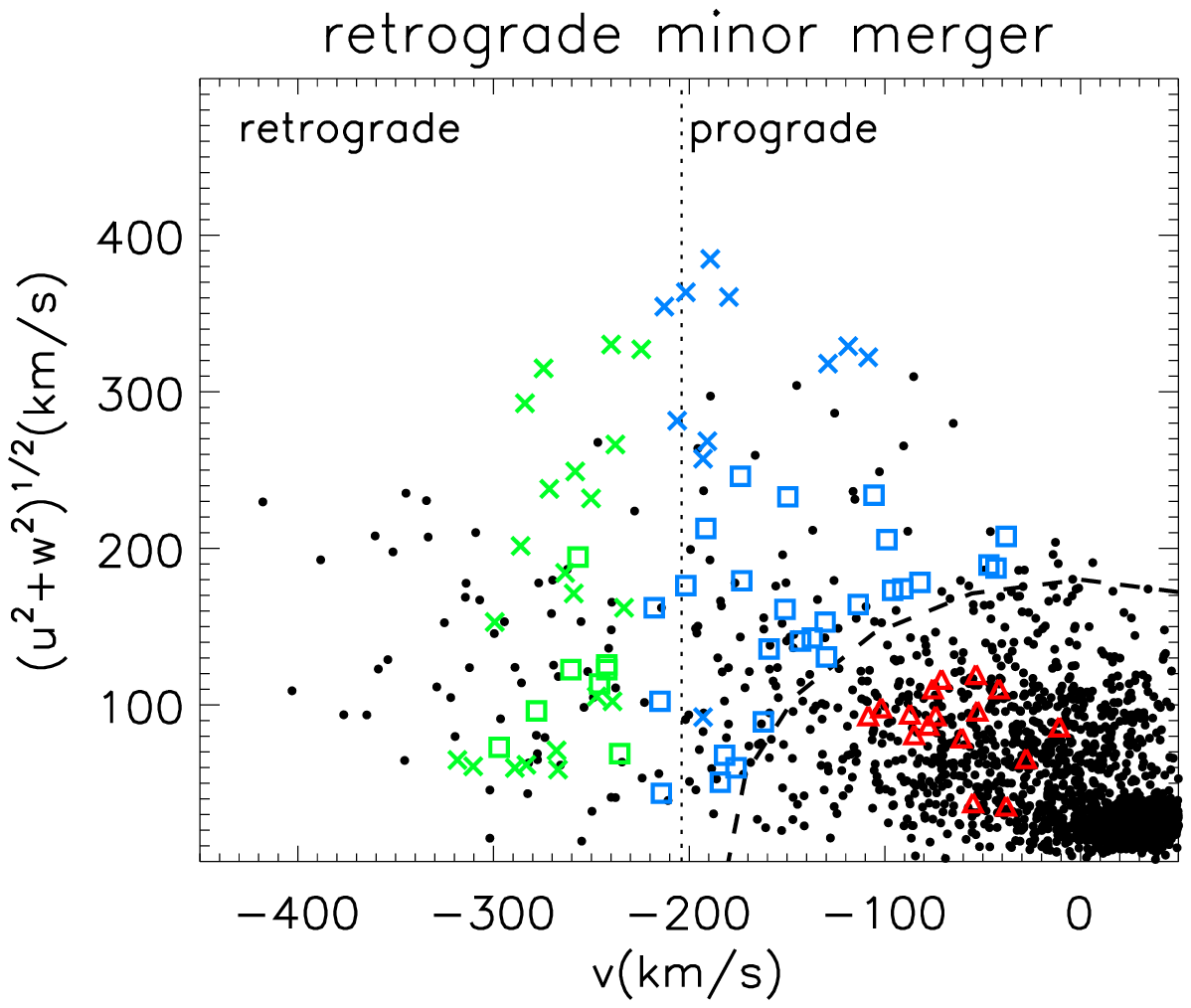}
\vspace{0cm}\hspace{1.2cm}
\includegraphics[width=4.8cm,angle=0]{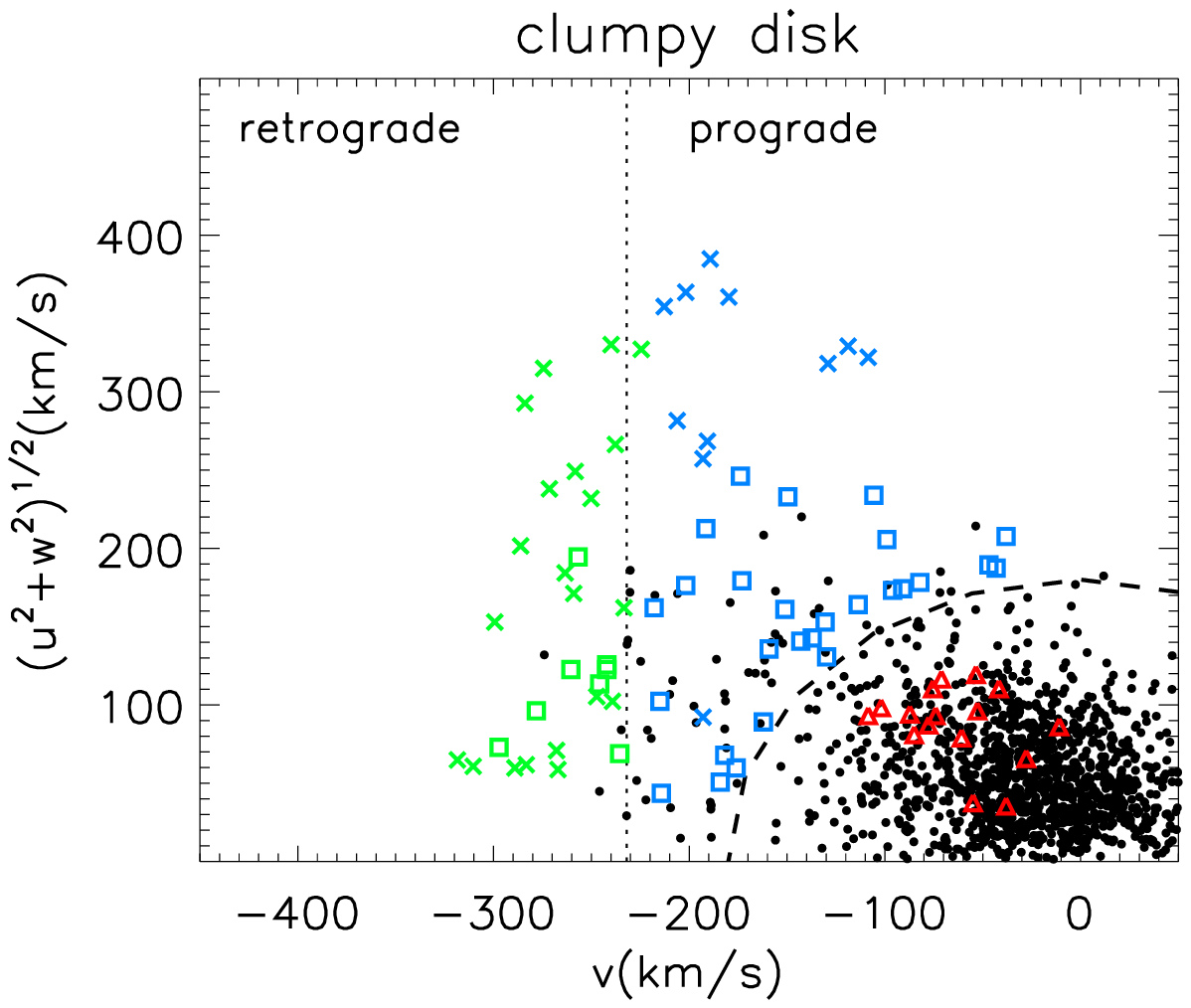}
\vspace{0.7cm} \caption{Toomre diagrams of stars at 2-3$r_d$
after a prograde minor merger (left), a retrograde minor merger
(middle) and in a clumpy disk evolved in isolation for 3~Gyr. Plotted are
the quadrature sum of vertical and radial velocities ($u$ and $w$) as
function of the rotational velocity, $v$, relative to the Local Standard of
Rest. The primary galaxy in the merger models has a gas-to-stellar mass ratio of
$f_{gas}=0.2$. The dashed line marks a total velocity of 180~km s$^{-1}$,
which serves as a potential criteria to separate thick disk
stars and halo stars \citep{venn204}, and the dotted line is the Local
Standard of Rest in these models. Also shown are the observed thick disk stars (triangles), halo
stars with high $\alpha/Fe$ abundance (squares) and low
$\alpha/Fe$ value (crosses) in the solar neighborhood, see \citet{nissenS210}
for further details. The halo stars in retrograde motion are shown
in green whereas those in prograde motion are in blue.}
\label{toomre}
\end{figure*}

To what extent do the properties of our modeled thick disks formed
through minor mergers agree with observations? As previously discussed by
\cite{purcell210} and \cite{zolotov210}, minor mergers are able to considerably
heat a pre-existing thin disk. We have shown (\S~\ref{s-chars}) that the resulting
stellar vertical profile is characterized by a thick disk component and a stellar excess,
both of which can be fitted by $sech$ profiles with different scale
heights. The excess is a natural characteristic of minor merger models, which
cannot be reproduced by secular mechanisms, not even through scattering by
clumps in unstable gas-rich disks. Identifying these excess stars and
determining their properties
may thus provide a way to constrain thick disk formation scenarios,
but they may be confused with stars in the inner halo and be
difficult to find and study in practice \citep{stanway08, carollo10}.

Interestingly, some of the excess stars may already have been
observed. \cite{nissenS210} reported the presence in the solar
neighborhood of stars with halo kinematics, but $\alpha/Fe$ abundances
similar to those of thick disk stars. These stars may well be part of the
stellar excess component, as we have shown (\S~\ref{kinem}) that excess stars lag rotationally with
respect to thick disk stars. If the Milky Way thick disk formed through
the heating of a pre-existing thin component by minor mergers at early
epochs, we should expect to find an inner halo population (the excess),
with abundances similar to those of stars in the thick disk, but lagging
with respect to them. Compared to stars originally in the thin disk,
stars accreted from satellite galaxies may have abundances which are lower
(though probably higher than those of dwarf galaxies
currently orbiting in the Milky Way halo) and probably also a lower
angular momentum content \citep[see][]{villalobosKH210}.
In summary, in the merger scenario the inner halo stars
would probably have a population with abundances similar to those of the
thick disk, lagging with respect to thick disk stars, but rotating more
rapidly than accreted satellites.

The Toomre diagram shows the relationship between the total radial and
vertical kinetic energy of stars and the rotational energy relative to
the Local Standard of Rest \citep{sandageF187}. As it is based on
kinematic energy, it provides us with an appropriate tool to evaluate
the results of our models, which are kinematic in nature, and a way to
compare with the kinematic properties of various components in the
Milky Way.

In minor merger models the stellar excess stars
lie exactly in the area of the Toomre diagram where $\alpha$ enhanced
stars are found \citep{nissenS210}, thus strengthening the hypothesis
that they may originally have been in the thin disk. Of course,
the dynamical and chemical properties of thick disks and halos will depend
on the properties of the satellite galaxy. We expect the scatter in the
dynamics and metallicity to increase with vertical distance, as the
fraction of satellite stars increases with height. On the other hand,
the clumpy disk model does not reproduce the Toomre diagram of the
Milky Way (see the right panel of Fig.~\ref{toomre}), since it cannot
heat disk stars as efficiently in the vertical direction as minor
mergers do.  It should be pointed out that our models agree with the
Toomre diagram of the Milky Way, although they are neither tuned to
reproduce the Milky Way galaxy, nor meant to capture the complex
accretion history that it may have had.

\begin{figure}
\vspace{0.5cm}
\centering
\includegraphics[width=5.3cm,angle=0]{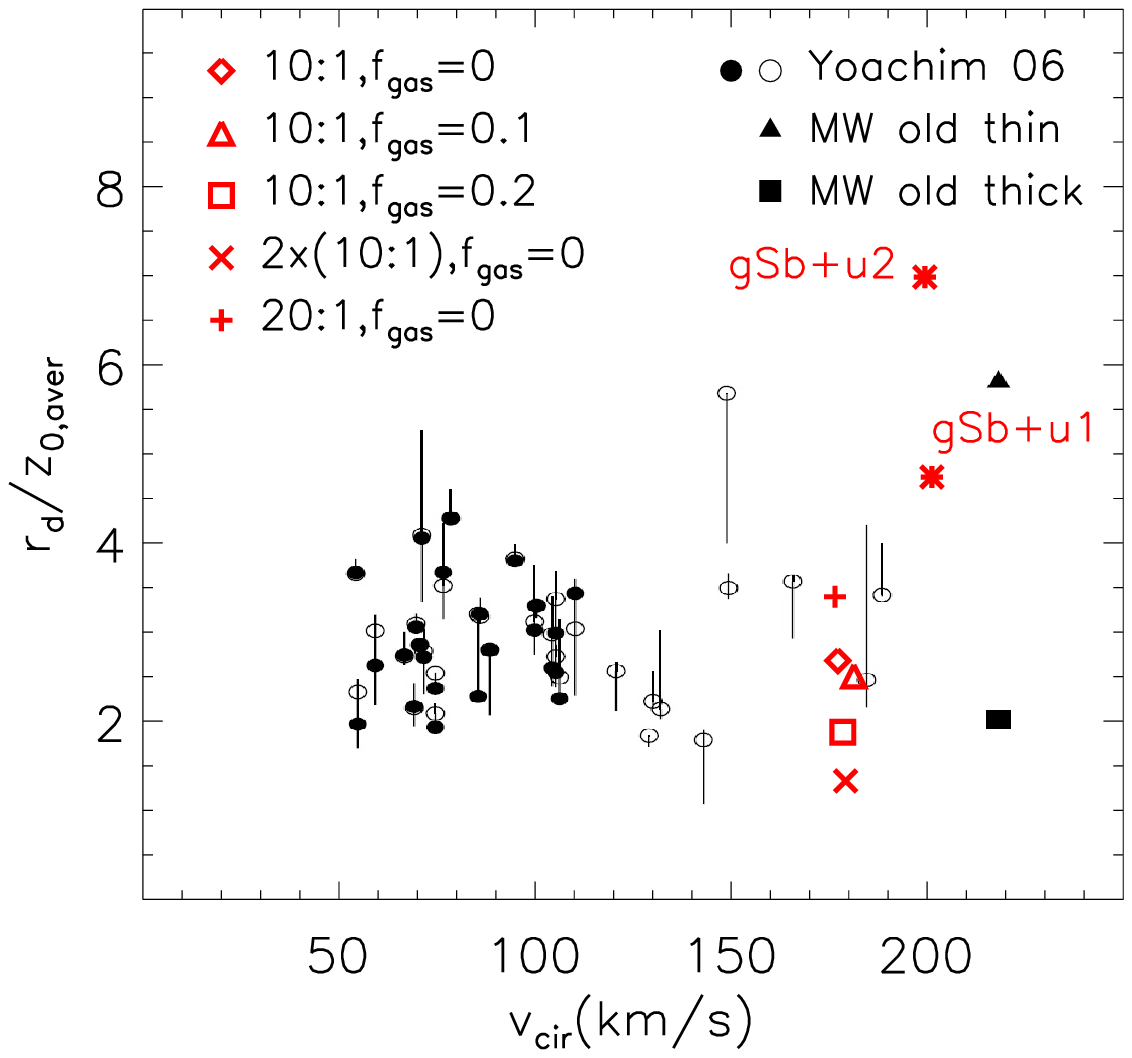}
\vspace{1.3cm} \caption{Ratio of the thick disk scale length and
scale height as function of circular velocity, $v_{cir}$, for five of our
dissipative/dissipationless and single/double minor merger models, as
well as for two unstable gas-rich clump disk models (gSb+u1 and gSb+u2, see
Fig.~\ref{thick-gas} for the notations). Also shown are observations
of a sample of late-type galaxies from \cite{yoachimD205} (open and
filled circles) and of the thin and old thick disk of the Milky Way
\citep{larsenH203}.}
\label{ratio}
\end{figure}

Adding a dissipative component (gas) decreases both the vertical
scale height and the radial scale length of the thick disk created
during a minor merger (see \S~\ref{gas}). For a gas-to-stellar mass
fraction $f_{gas}=0.2$, the decrease in thick disk scale height is
about $\sim$20$\%$ with respect to the dissipationless case, and the
decrease in scale length is up to 15$\%$-20$\%$. The ratio of thick
disk scale length and particle-number-averaged scale height
($r_{d}/z_{0,aver}$) is shown in Fig.\ref{ratio} as function of the
circular velocity of the disk, $v_{cir}$, for simulations with a
variety of mass ratio, gas fraction and merger histories. Here
$v_{cir}$ is estimated at 4 times the disk scale length, $r_{d}$,
where the disk rotation curve becomes flat, and $z_{0,aver}$ is the
number-of-particles-weighted average thick disk scale height within
$r<4r_{d}$. Since all our primary galaxies have the same mass, as we
are basically concerned with primary:satellite mass ratios in our
merger simulations, we only sample a single circular velocity.
However, we note that more or less massive primaries would produce
similar results. We compare our results with observations of
late-type galaxies and the MW \citep{larsenH203, yoachimD205}. For
all our minor merger models, whether gas-free or with gas, or single
or double merger events, the ratios are consistent with those of
extragalactic thick disks and the Milky Way old thick disk. Without
a gas component, two consecutive mergers effectively halve the ratio
by doubling the disk scale height and keeping the radial scale
length almost unchanged. The impact of accreting additional gas on
the scale height of the thick disk (which is likely to lead to its
contraction) would likely lower the $r_{d}/z_{0,aver}$ ratio,
irrespective of whether this additional gas is accreted by merging
of gas rich satellites or through accreting gas from the halo or
intergalactic medium.

Also shown in Fig.~\ref{ratio} are the ratios for two thick disks formed
through internal clumps scattering, which are radially more extended but
less thick. Their ratios are on average about 2 times larger than those
of the observed thick disks. In this case, additional gas accretion
from the galaxy halo or surroundings would only increase the discrepancy
between observations and the models.

\begin{figure}
\vspace{0.5cm}
\centering
\includegraphics[width=5.7cm,angle=0]{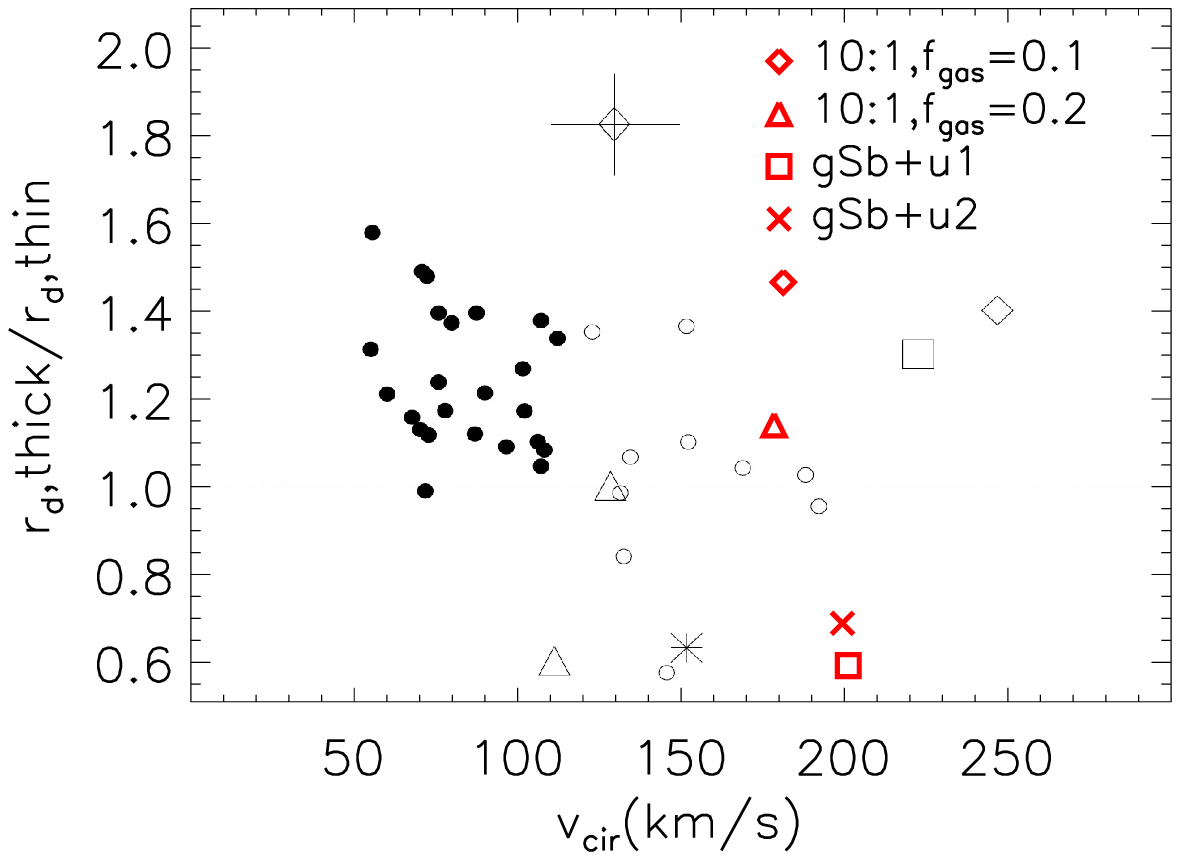}
\vspace{1.2cm} \caption{Comparison between observations and
our model results of thick-to-thin disk scale length ratios,
$r_{d,thick}/r_{d,thin}$, as function of circular velocity,
$v_{cir}$, from our minor merger models with two gas fractions,
$f_{gas}=0.1$ and 0.2, and two clumpy gas-rich disk models (gSb+u1 and
gSb+u2, see \S~\ref{model} for the notations). Observations
are shown for samples of late-type spirals \citep[solid
black and open circles]{yoachimD206}, \citep[][diamond and
triangles]{wu02,abeBC199,neeserSMP202}, a sample of early
type galaxies \citep[][diamond with a cross]{pohlen04} and the Milky Way
\citep[][square]{larsenH203}. In addition, a cosmological simulation
is shown \citep[from][asterisk]{brookKGF204}.}
\label{scalelengthsratio}
\end{figure}

The loss of angular momentum in the gas during the merger causes both the
deepening of the gravitational potential and enhanced star-formation in
the densest regions of the disk, which is expected to lead to a larger
scale length of the thick disk than of the thin disk newly formed during
the merger. This is also supported by observations \citep{pohlen07}. In
order to estimate the scale length of the thin disk, we consider only
stars at disk height $|z|<1$ kpc, whereas stars at greater heights
($1<|z|<5$ kpc) are considered to constitute the thick disk --
as was done in the analysis of observations of external galaxies
\citep[e.g.,][]{neeserSMP202}. A comparison between our results and
observations (Fig.~\ref{scalelengthsratio}) shows a very good overall
agreement in the ratio of thick to thin disk scale lengths. This is a natural
outcome of minor mergers with a primary disk containing gas and does
not require any fine tuning.

\citet{stewartBW208}, in a study of the merger history of dark matter
halos, found that virtually every MW-sized halo has experienced at
least one minor merger with mass ratio $\lesssim$10:1 in the last
10~Gyr \citep[see also][]{guo08}. One minor merger of mass ratio
$\lesssim$10:1, as we have shown, is not going to destroy the disk
and leads to something that looks like a thin+thick disk. However,
the effect of dissipationless minor mergers on thickening the
stellar disk is cumulative. If the merger frequency is higher, then
the merger-induced disk thickening will efficiently destroy the
original thin disk, leading to a spheroid dominated early type
galaxy. This predicament implies that the evolution history of most
disk galaxies must be relatively quiet and suggests that there is a
limit to how much mass can be accreted solely through minor mergers.
However, we have yet to investigate dissipative multiple mergers and
expect that the addition of gas will allow some additional
dissipation of the energy and angular momentum of the merger. This
would also have the effect of erasing the perhaps unique signatures
of the timing of the previous event and result in a more complex
evolution \citep[see also][]{brookGMK205}.

The question arises if metallicity or metallicity gradients can be used
to distinguish between various thick disk formation models. Within a
cosmological context, \citet{brookGMK205} discussed that merger-induced
star-formation and on-going gas accretion could result in a thick disk
stellar population with a homogeneous age and metallicity distribution. In
local edge-on galaxies \citep{mould205,sethDdJ205,rejkuba09},
no-significant vertical color gradients are observed, which suggests
the absence of metallicity or age gradients.  Although some disk
galaxies in the local universe show strong radial metallicity gradients
\citep[e.g.,][]{shields90}, they do not constitute a stringent test of the merger
hypothesis, given the dependence on the pre-existing disk and the strong
mixing expected.

Our models would not necessarily predict a strong radial or vertical
gradient in metallicity or age, for two main reasons: the first is that
a vast majority of the stars in the thick disk at heights $\lesssim$4-5~kpc
are dominated by stars originally in the thin disk, which implies
that any radial or vertical gradients in either age or metallicity depend on the
properties of the pre-existing disk and not on the properties of the
merging satellite galaxy. The second is that mergers are effective in radial
mixing, as the induced changes in the orbital families of the
disk and in the velocity dispersion would diminish any intrinsic gradients
that the original disk may have had \citep{haywood08,schoenrich09}.

Furthermore, accreting gas from either the surrounding dark matter halo or cosmic
filaments will lower the disk scale height over time. This would have
the impact of increasing the ratio of the scale lengths of the thin and
thick disks. We have found good agreement for this ratio between
our simulations and observations of nearby galaxies
(Fig.~\ref{ratio}). The agreement would still be reasonable if the
scale height of the thick disk were to decrease by almost a factor of
2 (through an accretion-induced increase in the mass surface density
of a factor of 2) in our minor merger simulations. For the unstable
disk models, which already produce a ratio that is too large, a
decrease in the scale height would make the discrepancy even worse.

However, we have yet to investigate multiple mergers which are
separated by episodes of significant gas accretion from the halo or
filaments. We would expect that such gas will allow for additional
dissipation of energy and angular momentum of any subsequent mergers
and make the disk more robust against destruction. Thus if gas
accretion is important for galaxy growth, the limit on the number of
minor mergers before the effective destruction of the disk may be
larger than indicated by the results presented here.

The caveat is of course that the variance in the metallicity of the thick disk stars will have increased if
the stellar population of the disk has evolved significantly in between
merger events. Of course, if gas is accreted and depending on its metallicity and how much
star-formation occurs, the metallicity of the disk may have increased
or perhaps even decreased. Therefore, for an inherently stochastic
process like the minor merger-induced thick disk formation scenario,
the key factor for determining the characteristics of the mergers is
not the mean value of the stellar metallicity, but its variance in
abundances \citep{richard10}.

As observations indicate that thick disks
are clearly distinct from the thin disk component
\citep[e.g.,][]{prochaska00,feltzing03,bensby104,bensby204,bensby06,
fuhrmann08} and that they must have formed early and relatively fast,
this implies that the only viable formation scenarios are those that
form thick disks early, with the thin and the thick disk aligned --
with either a parallel or a perpendicular angular momentum vector --
and with a generally relatively larger scale length for the thick disks
\citep{pohlen07}. The minor merger-induced thick disk formation scenario
merely requires a pre-existing disk in order to work. In all other
scenarios the thick disk either relies critically on the pre-existing disk
itself, with the thick and the thin disk at least partially co-evolving --
as in the secular evolution models, which must also explain lagging and
counter-rotating thick disks -- or the thick disk predates the thin disk
formation. The difference in scale lengths of the thick and thin disks
and the weak chemical and dynamical relationship between the two suggest
that any viable mechanism should preserve their rough alignment and have
much of the mass of the thin disk form later.  Clearly, an early phase of
multiple minor mergers would be consistent with the data if such a phase
were to last sufficiently long to allow for some enrichment due to type Ia supernovae.
Issues of timing and total accreted mass are also important, so as not to
destroy the disk completely, which would then provide a natural way
for the angular momentum vectors to remain parallel over a Hubble time.

Because mergers can cause strong gas inflows \citep[e.g.,][]{montuori210},
we would also expect that the properties of the merger-induced thick
disk relate somehow to the bulge growth. Although not discussed in detail in this study,
we indeed find that the bulge component does grow during minor
mergers, thus slowly shifting the galaxies towards earlier Hubble
types, and that the effects of minor mergers are cumulative in thickening
the disk. A natural outcome of minor mergers is therefore that both the significance and scale height
of the thick disk should be related to whether a galaxy is a late- or
early-type disk type. Moreover, since mergers can have a wide range of
initial orbital parameters, we would expect to find counter-rotating
thick disk stars if the interaction are violent enough or that the
some satellites haves sufficiently high density to begin to contribute
significantly to the kinematics of the thick disk \citep{quDM211}.
Observations have suggested that the thick disk scale
height is related to Hubble type \citep{degrijs197}, that some disk
dominated late-type galaxies do not have significant thick disks, and
that some thick disks counter-rotate \citep{morrisonBH194,
  gilmoreWN202, yoachimD205}.

Finally, our galaxy models show properties typical of local disk
galaxies. Although the details are controversial, if the MW formed
both its halo and thick disk at around the same time, over 10~Gyr
ago \citep{prochaska00,feltzing03, bensby104,bensby204,fuhrmann04},
then the initial conditions of the MW disk would have been probably
different from what we assumed. However, due to the lack of knowledge of the properties
of disks in the early universe, the initial conditions adopted for
our galaxy models represent a reasonable starting point for an exploration
of the impact of gas fraction, orbital parameters and merging history
on thick disk formation.

\section{Conclusions}\label{conclu}

We used N-body numerical simulations to investigate
if minor mergers are a plausible mechanism to form thick disks that
have morphological and kinematic properties consistent with the observations, in
particular the vertical surface density profile and the disk scale
height. We explored in particular the stellar excess formed through
minor mergers. The possibility that minor mergers can scatter stars
originally in the thin disk (or in the bulge) to large heights from
the galaxy mid-plane has recently been pointed out by \cite{purcell210}
and \cite{zolotov210}, but we have for the first time systematically studied
the properties of this excess: its vertical distribution, its dependence on gas fraction,
orbital parameters and consecutive mergers. Our minor merger simulations cover a variety of
initial configuration parameters, e.g. gas-to-stellar mass fractions
($f_{gas}=0$, 0.1, and 0.2), initial orbital energies and orientations,
two different mass ratio (20:1 and 10:1), as well as the cumulative
effect of two consecutive dissipationless minor mergers.

The main results of our study can be summarized as follows:

\begin{itemize}

\item Minor interactions efficiently and gradually start heating
  pre-existing thin stellar disks from the first pericenter passage of
  the satellite galaxy, well before merging is complete. The resulting
  vertical surface density profile of the stellar thick disk is
  characterized by a sech function. Furthermore, a stellar excess is
  visible in the profile at greater heights $z\gtrsim 4-5$~kpc in the
  inner regions ($r\lesssim 3R_{d}$), which can also be fitted by an
  additional sech function with a relatively larger scale height. Due
  to this excess, a double-sech profile is necessary to fit the entire
  vertical surface density profile;

\item Usually the vertical stellar profile is dominated at all disk
  heights by stars that were initially in the disk of the primary
  galaxy. Only in a few dissipationless merger models we find that the
  stellar excess at greater heights is dominated by the contribution
  of satellite stars;

\item The excess has morphological and kinematical properties which
  are distinct generally, but see the 2$\times$10:1 case in Fig.~\ref{dens-fit-multi} from
  those of thick disk stars (however, see our most massive accretion
  simulation in Fig.~\ref{dens-fit-multi}, the 2$\times$10:1 case);

\item The scale height of the stellar excess is constant in radius,
  while the thick disk scale height increases with radius -- this disk
  flaring is a characteristic of thick disks formed by minor mergers,
  as shown previously by \cite{bournaudEM209};

\item The scale height of the thick disk decreases with an increase in
  the gas-to-stellar mass fraction of the primary disk, while that of
  the stellar excess does not: no significant dependence on the gas
  fraction in the primary disk is found, at least for $f_{gas}\le
  0.2$;

\item The effect of two consecutive mergers is cumulative, both on thick disk and stellar excess scale height. 
However, obviously, after some number of consecutive mergers, their impact will saturate,
but the exact number necessary for this will depend on a variety of
factors such as mass ratio, orbital parameters, etc, which has yet to be
explored;

\item Stars in the stellar excess rotate slower than stars in the
  thick disk, and their kinematics are compatible with those of
  high-$\alpha$ abundant stars found recently by \cite{nissenS210} in
  the solar neighborhood;

\item In minor mergers the scale length of the thick disk is larger
  than that of the thin disk, consistent with observations of the
  Milky Way and other nearby galaxies. In simulations with higher (or
  lower) gas fractions the ratio of thick to thin disk scale lengths
  is lower (or higher). Whereas ratios larger than one are a natural
  outcome of minor mergers, the evolution of a ''clumpy disk''
  (simulating, at least to some extent, the influence of
secular processes on the evolution of
  the disk) produces small ratios that are generally inconsistent with
  observations. Subsequent gas accretion from the halo, or along
  filaments in the inter-galactic medium, will likely decrease the
  thick disk scale height. While this does not represent a major
  problem for the minor merger scenario, for the unstable disk model
  it makes its discrepancy with observations worse.

\end{itemize}

We also compared the properties of thick disks formed by minor mergers and through
strong scattering by self-gravitating clumps in unstable gas rich disks
(a secular ``clumpy disk'' scenario). We find that the clumpy disks have
smaller disk scale heights, $z_{0}\approx 0.85$~kpc, which are independent
of radius \citep[in agreement with the results of][]{bournaudEM209} and
unlike the merger-driven thick disks. The thick disks
formed through scattering do not show any stellar excess at large disk
scale heights, unlike the merger simulations. Thus a possible way to distinguish between this secular
thick disk formation scenario and minor merger models is to investigate
the radial distribution of scale heights, as well as the presence and
properties of the excess component in the vertical stellar profile at large disk heights.

Our results also put limits on the amount of accretion that thin disks
can undergo in their lifetimes. This is not a simple function of accreted
mass, as merger-induced star formation within the pre-existing disk will
act to both stabilize the disk against destruction and to limit the
mass fraction of thick disk stars by dissipating part of the energy of
the interaction. Another component which may play an important role in
thick disk formation is the bulge component, as observations have implied
that thick disk properties relate to the Hubble type. The existence of a
pre-merger bulge component, or its growth during the merger process, could prevent the
stellar disk from being tidally distorted, or weaken the disk heating,
and thus result in a relatively less thick stellar disk in the post-merger phase.
Detailed study of the interplay between galactic bulge growth and thick
disk formation could further our knowledge of the thick disk component
as well as galaxy evolution in general.

\section*{Acknowledgments}
YQ and PDM are supported by a grant from the French Agence Nationale
de la Recherche (ANR). We are grateful to Beno\^it Semelin and
Fran\c{c}oise Combes for developing the code used in this paper and
for their permission to use it. We thank the anonymous referee
for his/her useful and insightful comments which improved the paper
substantially. We also wish to thank Chanda Jog and Gary Mamon for
their useful comments. These simulations are available as part of
the GalMer simulation data base (\emph{http://galmer.obspm.fr}).

\bibliographystyle{aa}
\bibliography{thickdisk}

\begin{thebibliography}{92}
\expandafter\ifx\csname natexlab\endcsname\relax\def\natexlab#1{#1}\fi

\bibitem[{Abadi {et~al.}(2003)Abadi, Navarro, Steinmetz, \& Eke}]{abadiNSE203}
Abadi, M.~G., Navarro, J.~F., Steinmetz, M., \& Eke, V.~R. 2003, ApJ, 597, 21

\bibitem[{Abe {et~al.}(1999)Abe, Bond, Carter, Dodd, Fujimoto, Hearnshaw,
  Honda, Jugaku, Kabe, Kilmartin, \& coauthors}]{abeBC199}
Abe, F., Bond, I.~A., Carter, B.~S., {et~al.} 1999, AJ, 118, 261

\bibitem[{Aguerri {et~al.}(2001)Aguerri, Balcells, \& Peletier}]{aguerri01}
Aguerri, J. A.~L., Balcells, M., \& Peletier, R.~F. 2001, A\&A, 367, 428

\bibitem[{Banerjee \& Jog(2007)}]{banerjeeJ207}
Banerjee, A. \& Jog, C.~J. 2007, ApJ, 662, 335

\bibitem[{Barbanis \& Woltjer(1967)}]{barbanisW167}
Barbanis, B. \& Woltjer, L. 1967, ApJ, 150, 461

\bibitem[{{Bensby} \& {Feltzing}(2006)}]{bensby06}
{Bensby}, T. \& {Feltzing}, S. 2006, MNRAS, 367, 1181

\bibitem[{{Bensby} {et~al.}(2004{\natexlab{a}}){Bensby}, {Feltzing}, \&
  {Lundstr{\"o}m}}]{bensby104}
{Bensby}, T., {Feltzing}, S., \& {Lundstr{\"o}m}, I. 2004{\natexlab{a}}, A\&A,
  415, 155

\bibitem[{{Bensby} {et~al.}(2004{\natexlab{b}}){Bensby}, {Feltzing}, \&
  {Lundstr{\"o}m}}]{bensby204}
{Bensby}, T., {Feltzing}, S., \& {Lundstr{\"o}m}, I. 2004{\natexlab{b}}, A\&A,
  421, 969

\bibitem[{Binney \& Tremaine(1987)}]{binneyT187}
Binney, J. \& Tremaine, S. 1987, Galacitc Dynamics (Princeton: princeton Univ.
  Press)

\bibitem[{Bournaud {et~al.}(2009)Bournaud, Elmegreen, \&
  Martig}]{bournaudEM209}
Bournaud, F., Elmegreen, B.~G., \& Martig, M. 2009, ApJ, 707, L1

\bibitem[{Brook {et~al.}(2007)Brook, Richard, Kawata, Martel, \&
  Gibson}]{brookRKMG207}
Brook, C., Richard, S., Kawata, D., Martel, H., \& Gibson, B.~K. 2007, ApJ,
  658, 60

\bibitem[{Brook {et~al.}(2005)Brook, Gibson, Martel, \& Kawata}]{brookGMK205}
Brook, C.~B., Gibson, B.~K., Martel, H., \& Kawata, D. 2005, ApJ, 630, 298

\bibitem[{Brook {et~al.}(2004)Brook, Kawata, Gibson, \& Freeman}]{brookKGF204}
Brook, C.~B., Kawata, D., Gibson, B.~K., \& Freeman, K.~C. 2004, ApJ, 612, 894

\bibitem[{Burstein(1979)}]{burstein179}
Burstein, D. 1979, ApJ, 234, 829

\bibitem[{Carlberg \& Sellwood(1985)}]{carlbergS185}
Carlberg, R.~G. \& Sellwood, J.~A. 1985, ApJ, 292, 79

\bibitem[{Carney {et~al.}(1989)Carney, Latham, \& Laird}]{carneyLL189}
Carney, B.~W., Latham, D.~W., \& Laird, J.~B. 1989, AJ, 97, 423

\bibitem[{{Carollo} {et~al.}(2010){Carollo}, {Beers}, {Chiba}, {Norris},
  {Freeman}, {Lee}, {Ivezi{\'c}}, {Rockosi}, \& {Yanny}}]{carollo10}
{Carollo}, D., {Beers}, T.~C., {Chiba}, M., {et~al.} 2010, ApJ, 712, 692

\bibitem[{Casetti-Dinescu {et~al.}(2010)Casetti-Dinescu, Girard, Korchagin, \&
  van Altena}]{casetti210}
Casetti-Dinescu, D.~I., Girard, T.~M., Korchagin, V.~I., \& van Altena, W.~F.
  2010, arXiv1011.6253

\bibitem[{Chiba \& Beers(2000)}]{chibaB200}
Chiba, M. \& Beers, T.~C. 2000, AJ, 119, 2843

\bibitem[{Chilingarian {et~al.}(2010)Chilingarian, Di~Matteo, Combes, Melchior,
  \& Semelin}]{chilingarianDMC209}
Chilingarian, I., Di~Matteo, P., Combes, F., Melchior, A.-L., \& Semelin, B.
  2010, A\&A, submitted; arXiv:1003.3243

\bibitem[{Comer\'{o}n {et~al.}(2011)Comer\'{o}n, Knapen, Sheth, Regan, Hinz,
  Gil~de Paz, Men\'{e}ndez-Delmestre, Mu\~{n}oz Mateos, Seibert, Kim, \&
  coauthors}]{comeron211}
Comer\'{o}n, S., Knapen, J.~H., Sheth, K., {et~al.} 2011, ApJ, 729, 18

\bibitem[{Dalcanton \& Bernstein(2002)}]{dalcantonB202}
Dalcanton, J.~J. \& Bernstein, R.~A. 2002, AJ, 124, 1328

\bibitem[{de~Grijs \& Peletier(1997)}]{degrijs197}
de~Grijs, R. \& Peletier, R.~F. 1997, A\&A, 320, 21

\bibitem[{Di~Matteo {et~al.}(2008)Di~Matteo, Bournaud, Martig, Combes,
  Melchior, \& Semelin}]{dimatteo08}
Di~Matteo, P., Bournaud, F., Martig, M., {et~al.} 2008, A\&A, 492, 31

\bibitem[{Di~Matteo {et~al.}(2010)Di~Matteo, Lehnert, Qu, \& van
  Driel}]{dimatteo210}
Di~Matteo, P., Lehnert, M.~D., Qu, Y., \& van Driel, W. 2010, A\& A, 525, L3

\bibitem[{Dierickx {et~al.}(2010)Dierickx, Klement, Rix, \& Liu}]{dierickx210}
Dierickx, M., Klement, R., Rix, H.-W., \& Liu, C. 2010, ApJ, 725, 186

\bibitem[{{Feltzing} {et~al.}(2003){Feltzing}, {Bensby}, \&
  {Lundstr{\"o}m}}]{feltzing03}
{Feltzing}, S., {Bensby}, T., \& {Lundstr{\"o}m}, I. 2003, A\&A, 397, L1

\bibitem[{Fry {et~al.}(1999)Fry, Morrison, Harding, \& Boroson}]{fry199}
Fry, A.~M., Morrison, H.~L., Harding, P., \& Boroson, T.~A. 1999, AJ, 118, 1209

\bibitem[{{Fuhrmann}(2004)}]{fuhrmann04}
{Fuhrmann}, K. 2004, AN, 325, 3

\bibitem[{{Fuhrmann}(2008)}]{fuhrmann08}
{Fuhrmann}, K. 2008, MNRAS, 384, 173

\bibitem[{Gilmore \& Reid(1983)}]{gilmoreR183}
Gilmore, G. \& Reid, N. 1983, MNRAS, 202, 1025

\bibitem[{Gilmore {et~al.}(2002)Gilmore, Wyse, \& Norris}]{gilmoreWN202}
Gilmore, G., Wyse, R. F.~G., \& Norris, J.~E. 2002, ApJ, 574, 39

\bibitem[{Guo \& White(2008)}]{guo08}
Guo, Q. \& White, S. D.~M. 2008, MNRAS, 384, 2

\bibitem[{{Haywood}(2008)}]{haywood08}
{Haywood}, M. 2008, MNRAS, 388, 1175

\bibitem[{Hernquist \& Quinn(1989)}]{hernquistQ189}
Hernquist, L. \& Quinn, P.~J. 1989, in The Epoch of Galaxy Formation, ed. C.S.
  Frenk, NATO ASI Series C, 264, 435

\bibitem[{Ivezi\'{c} {et~al.}(2008)Ivezi\'{c}, Sesar, Juri\'{c}, Bond,
  Dalcanton, Rockosi, Yanny, Newberg, Beers, Allende~Prieto, \&
  coauthors}]{ivezic208}
Ivezi\'{c}, Z., Sesar, B., Juri\'{c}, M., {et~al.} 2008, ApJ, 684, 287

\bibitem[{Kazantzidis {et~al.}(2008)Kazantzidis, Bullock, Zentner, Kravtsov, \&
  Moustakas}]{kazantzidis208}
Kazantzidis, S., Bullock, J.~S., Zentner, A.~R., Kravtsov, A.~V., \& Moustakas,
  L.~A. 2008, ApJ, 688, 254

\bibitem[{Lacey(1984)}]{lacey184}
Lacey, C.~G. 1984, MNRAS, 208, 687

\bibitem[{Larsen \& Humphreys(2003)}]{larsenH203}
Larsen, J.~A. \& Humphreys, R.~M. 2003, AJ, 125, 1958

\bibitem[{{Montuori} {et~al.}(2010){Montuori}, {Di Matteo}, {Lehnert},
  {Combes}, \& {Semelin}}]{montuori210}
{Montuori}, M., {Di Matteo}, P., {Lehnert}, M.~D., {Combes}, F., \& {Semelin},
  B. 2010, A\&A, 518, 56

\bibitem[{Morrison {et~al.}(1994)Morrison, Boroson, \& Harding}]{morrisonBH194}
Morrison, H.~L., Boroson, T.~A., \& Harding, P. 1994, AJ, 108, 1191

\bibitem[{Morrison {et~al.}(1997)Morrison, Miller, Harding, Stinebring, \&
  Boroson}]{morrisonMHS197}
Morrison, H.~L., Miller, E.~D., Harding, P., Stinebring, D.~R., \& Boroson,
  T.~A. 1997, AJ, 113, 2061

\bibitem[{Moster {et~al.}(2010)Moster, Macci\'{o}, Somerville, Johansson, \&
  Naab}]{mosterMS209}
Moster, B.~P., Macci\'{o}, A.~V., Somerville, R.~S., Johansson, P.~H., \& Naab,
  T. 2010, MNRAS, 403, 1009

\bibitem[{Mould(2005)}]{mould205}
Mould, J. 2005, AJ, 129, 698

\bibitem[{Narayan \& Jog(2002)}]{narayanj202b}
Narayan, C.~A. \& Jog, C.~J. 2002, A\&A, 390, L35

\bibitem[{N\"{a}slund \& J\"{o}rs\"{a}ter(1997)}]{naslundJ197}
N\"{a}slund, M. \& J\"{o}rs\"{a}ter, S. 1997, A\&A, 325, 915

\bibitem[{Neeser {et~al.}(2002)Neeser, Sackett, De~Marchi, \&
  Paresce}]{neeserSMP202}
Neeser, M.~J., Sackett, P.~D., De~Marchi, G., \& Paresce, F. 2002, A\&A, 383,
  472

\bibitem[{Nissen(1995)}]{nissen195}
Nissen, P.~E. 1995, IAUS, 164, 109

\bibitem[{Nissen \& Schuster(2010)}]{nissenS210}
Nissen, P.~E. \& Schuster, W.~J. 2010, A\&A, 511, 10

\bibitem[{Norris(1986)}]{norris186}
Norris, J. 1986, ApJS, 61, 667

\bibitem[{Parker {et~al.}(2004)Parker, Humphreys, \& Beers}]{parkerHB204}
Parker, J.~E., Humphreys, R.~M., \& Beers, T.~C. 2004, AJ, 127, 1567

\bibitem[{{Pohlen} {et~al.}(2004){Pohlen}, {Balcells}, {L{\"u}tticke}, \&
  {Dettmar}}]{pohlen04}
{Pohlen}, M., {Balcells}, M., {L{\"u}tticke}, R., \& {Dettmar}, R. 2004, A\&A,
  422, 465

\bibitem[{{Pohlen} {et~al.}(2007){Pohlen}, {Zaroubi}, {Peletier}, \&
  {Dettmar}}]{pohlen07}
{Pohlen}, M., {Zaroubi}, S., {Peletier}, R.~F., \& {Dettmar}, R. 2007, MNRAS,
  378, 594

\bibitem[{{Prochaska} {et~al.}(2000){Prochaska}, {Naumov}, {Carney},
  {McWilliam}, \& {Wolfe}}]{prochaska00}
{Prochaska}, J.~X., {Naumov}, S.~O., {Carney}, B.~W., {McWilliam}, A., \&
  {Wolfe}, A.~M. 2000, AJ, 120, 2513

\bibitem[{Purcell {et~al.}(2010)Purcell, Bullock, \& Kazantzidis}]{purcell210}
Purcell, C.~W., Bullock, J.~S., \& Kazantzidis, S. 2010, MNRAS, 404, 1711

\bibitem[{Qu {et~al.}(2010)Qu, Di~Matteo, Lehnert, van Driel, \& Jog}]{quDM210}
Qu, Y., Di~Matteo, P., Lehnert, M., van Driel, M., \& Jog, C.~J. 2010, A\&A,
  515, 11

\bibitem[{Qu {et~al.}(2011)Qu, Di~Matteo, Lehnert, van Driel, \& Jog}]{quDM211}
Qu, Y., Di~Matteo, P., Lehnert, M., van Driel, M., \& Jog, C.~J. 2011,
  submitted to A\&A

\bibitem[{Quinn {et~al.}(1993)Quinn, Hernquist, \& fullager}]{quinnhf193}
Quinn, P.~J., Hernquist, L., \& fullager, D.~P. 1993, ApJ, 403, 74

\bibitem[{Ratnatunga \& Freeman(1989)}]{ratnatungaF189}
Ratnatunga, K.~U. \& Freeman, K.~C. 1989, ApJ, 339, 126

\bibitem[{Rauscher {et~al.}(1998)Rauscher, Lloyd, Barnaby, Harper, Hereld,
  Loewenstein, Severson, \& Mrozek}]{rauscherLB198}
Rauscher, B.~J., Lloyd, J.~P., Barnaby, D., {et~al.} 1998, ApJ, 506, 116

\bibitem[{{Rejkuba} {et~al.}(2009){Rejkuba}, {Mouhcine}, \&
  {Ibata}}]{rejkuba09}
{Rejkuba}, M., {Mouhcine}, M., \& {Ibata}, R. 2009, MNRAS, 396, 1231

\bibitem[{{Richard} {et~al.}(2010){Richard}, {Brook}, {Martel}, {Kawata},
  {Gibson}, \& {Sanchez-Blazquez}}]{richard10}
{Richard}, S., {Brook}, C.~B., {Martel}, H., {et~al.} 2010, MNRAS, 402, 1489

\bibitem[{Sales {et~al.}(2009)Sales, Helmi, Abadi, Brook, G\'{o}mez,
  Ro\v{s}kar, Debattista, House, Steinmetz, \& Villalobos}]{salesHAB209}
Sales, L.~V., Helmi, A., Abadi, M.~G., {et~al.} 2009, MNRAS, 400, 61

\bibitem[{Sandage \& Fouts(1987)}]{sandageF187}
Sandage, A. \& Fouts, G. 1987, AJ, 93, 592

\bibitem[{{Sch{\"o}nrich} \& {Binney}(2009)}]{schoenrich09}
{Sch{\"o}nrich}, R. \& {Binney}, J. 2009, MNRAS, 396, 203

\bibitem[{Schwarzkopf \& Dettmar(2000)}]{schwarzkopfD200}
Schwarzkopf, U. \& Dettmar, R.-J. 2000, A\&A, 361, 451

\bibitem[{Semelin \& Combes(2002)}]{semelinC202}
Semelin, B. \& Combes, F. 2002, A\&A, 388, 826

\bibitem[{Seth {et~al.}(2005)Seth, Dalcanton, \& de~Jong}]{sethDdJ205}
Seth, A.~C., Dalcanton, J.~J., \& de~Jong, R.~S. 2005, AJ, 130, 1574

\bibitem[{Shaw \& Gilmore(1989)}]{shawG189}
Shaw, M.~A. \& Gilmore, G. 1989, MNRAS, 237, 903

\bibitem[{Shaw \& Gilmore(1990)}]{shawG190}
Shaw, M.~A. \& Gilmore, G. 1990, MNRAS, 242, 59

\bibitem[{{Shields}(1990)}]{shields90}
{Shields}, G.~A. 1990, ARA\&A, 28, 525

\bibitem[{Soubiran {et~al.}(2003)Soubiran, Bienaym\'e, \&
  Siebert}]{soubiranBS203}
Soubiran, C., Bienaym\'e, O., \& Siebert, A. 2003, A\&A, 398, 141

\bibitem[{Spitzer \& Schwarzschild(1951)}]{spitzerS151}
Spitzer, L.~J. \& Schwarzschild, M. 1951, ApJ, 114, 385

\bibitem[{{Stanway} {et~al.}(2008){Stanway}, {Bremer}, {Lehnert}, \&
  {Eldridge}}]{stanway08}
{Stanway}, E.~R., {Bremer}, M.~N., {Lehnert}, M.~D., \& {Eldridge}, J.~J. 2008,
  MNRAS, 384, 348

\bibitem[{Statler(1988)}]{statler188}
Statler, T.~S. 1988, ApJ, 331, 71

\bibitem[{Stewart {et~al.}(2008)Stewart, Bullock, Wechsler, Maller, \&
  Zentner}]{stewartBW208}
Stewart, K.~R., Bullock, J.~S., Wechsler, R.~H., Maller, A.~H., \& Zentner,
  A.~R. 2008, ApJ, 683, 597

\bibitem[{Tikhonov {et~al.}(2005)Tikhonov, Galazutdinova, \&
  Drozdovsky}]{tikhonovGD205}
Tikhonov, N.~A., Galazutdinova, O.~A., \& Drozdovsky, I.~O. 2005, A\&A, 431,
  127

\bibitem[{Tsikoudi(1979)}]{tsikoudi179}
Tsikoudi, V. 1979, ApJ, 234, 842

\bibitem[{van~der Kruit(1988)}]{vdkruit188}
van~der Kruit, P.~C. 1988, A\&A, 192, 117

\bibitem[{van~der Kruit \& Searle(1981{\natexlab{a}})}]{vdkruitS181a}
van~der Kruit, P.~C. \& Searle, L. 1981{\natexlab{a}}, A\&A, 95, 105

\bibitem[{van~der Kruit \& Searle(1981{\natexlab{b}})}]{vdkruitS181b}
van~der Kruit, P.~C. \& Searle, L. 1981{\natexlab{b}}, A\&A, 95, 116

\bibitem[{Venn {et~al.}(2004)Venn, Irwin, Shetrone, \& et~al.}]{venn204}
Venn, K.~A., Irwin, M., Shetrone, M.~D., \& et~al. 2004, AJ, 128, 1177

\bibitem[{Villalobos \& Helmi(2008)}]{villalobosH08}
Villalobos, A. \& Helmi, A. 2008, MNRAS, 391, 1806

\bibitem[{Villalobos \& Helmi(2009)}]{villalobosH09}
Villalobos, A. \& Helmi, A. 2009, MNRAS, 399, 166

\bibitem[{{Villalobos} {et~al.}(2010){Villalobos}, {Kazantzidis}, \&
  {Helmi}}]{villalobosKH210}
{Villalobos}, {\'A}., {Kazantzidis}, S., \& {Helmi}, A. 2010, apj, 718, 314

\bibitem[{Walker {et~al.}(1996)Walker, Mihos, \& Hernquist}]{walker196}
Walker, I.~R., Mihos, J.~C., \& Hernquist, L. 1996, ApJ, 460, 121

\bibitem[{Wilson {et~al.}(2010)Wilson, Helmi, Morrison, Breddels, Bienayme,
  Binney, Bland-Hawthorn, Campbell, Freeman, Fulbright, \&
  coauthors}]{wilson210}
Wilson, M., Helmi, A., Morrison, H.~L., {et~al.} 2010, arXiv1009.2052

\bibitem[{{Wu} {et~al.}(2002){Wu}, {Burstein}, {Deng}, {Zhou}, {Shang},
  {Zheng}, {Chen}, {Su}, {Windhorst}, {Chen}, {Zou}, {Xia}, {Jiang}, {Ma},
  {Xue}, {Zhu}, {Cheng}, {Byun}, {Chen}, {Deng}, {Fan}, {Fang}, {Kong}, {Li},
  {Lin}, {Lu}, {Sun}, {Tsay}, {Xu}, {Yan}, {Zhao}, \& {Zheng}}]{wu02}
{Wu}, H., {Burstein}, D., {Deng}, Z., {et~al.} 2002, AJ, 123, 1364

\bibitem[{Wyse \& Gilmore(1986)}]{wyseG186}
Wyse, R. F.~G. \& Gilmore, G. 1986, AJ, 91, 855

\bibitem[{Yoachim \& Dalcanton(2005)}]{yoachimD205}
Yoachim, P. \& Dalcanton, J. 2005, ApJ, 624, 701

\bibitem[{Yoachim \& Dalcanton(2006)}]{yoachimD206}
Yoachim, P. \& Dalcanton, J.~J. 2006, AJ, 131, 226

\bibitem[{Zolotov {et~al.}(2010)Zolotov, Willman, Brooks, Governato, Hogg,
  Shen, \& Wadsley}]{zolotov210}
Zolotov, A., Willman, B., Brooks, A.~M., {et~al.} 2010, ApJ, 721, 738

\end{thebibliography}
\end{document}